\documentclass[aps,prd,a4paper]{revtex4}
\begin{document}


\title{Nuclear spin structure in dark matter search: \protect\\
        The zero momentum transfer limit} 

\author{V.A. Bednyakov$^{a)}$ and F. \v Simkovic$^{b)}$}

\address{$^{a)}$Joint Institute for Nuclear Research, 
              Dzhelepov Laboratory of Nuclear Problems,
              141980 Dubna, Moscow Region, Russia \protect\\  
         $^{b)}$Department of Nuclear Physics, Comenius University,
              Mlynsk\'a dolina F1, SK--842 15 Bratislava, Slovakia}  

 
\vskip.5cm
 
\begin{abstract}
        We review the calculation of spin-dependent matrix elements 
        relevant to 
	scattering of weakly interacting massive 
        particles (WIMP) on nuclei. A comprehensive list, to
        our knowledge,  of the proton 
        and neutron total spin expectation values  
        ($\langle {\bf S}_{p} \rangle$ 
        and $\langle {\bf S}_{n} \rangle$) 
        calculated within different nuclear models is presented.
	These values allow a conclusion about the event rate expected 
	in direct dark matter search experiments  
	due to spin-dependent neutralino-nucleon interaction,
	provided neutralino is a dark matter particle.

\vskip 0.3cm
 
 
\end{abstract}
\maketitle

\section{Introduction}
	Historically, the spin-1/2 weakly interacting massive particles
        (WIMP) were considered as the first cold dark matter (DM) candidates.
	They interact with ordinary matter predominantly 
        by means of axial vector (spin-dependent) and 
        vector (spin-independent) couplings. 
 
	Nowadays, the main effort in the direct dark matter 
	search experiments is concentrated on the study of the 
	spin-independent (or scalar) interaction of 
	the dark matter particles with nuclei. 
	It is due to a strong (proportional to the
	squared mass of the target nucleus) 
	coherent enhancement of the dark matter 
	particle scalar interaction with nuclei.
	The results obtained in the field are
	presented in the form of the exclusion curves for
        the total even rate as a function of the mass of the dark matter
        particles. 
	The values of the cross section
	associated with the elastic scattering of  WIMP 
        due to scalar-nucleon interaction, which lie above these
        curves, are excluded.  
	There is also the so-called DAMA contour which corresponds to
	the first claim for evidence of the dark matter signal
\cite{Bernabei:2003za}.   

	The main goal of this review is to
	attract attention back to the 
	spin-dependent (or axial-vector) interaction of 
	dark matter particles with nuclei. The importance of this
        type of interaction of the DM particles is due to the reasons
        as follows: 
	i) the spin-dependent interaction  of the DM 
        particles provides us with twice stronger constraints on the SUSY 
        parameter space in comparison with the spin-independent interaction. 
        ii) in the case of spin-dependent interaction of heavy WIMPs
        with heavy target nuclei 
	the so-called long $q$-tail behavior of the relevant form--factor 
        allows detection of large nuclear recoil energy due to some
        nuclear structure effects;
        iii) it is worthwhile to note that by relying only upon
        the scalar interaction of the DM particles, which seems to
        be strongly suppressed, one might miss a DM signal
\cite{Bednyakov:2000he}. However, by a simultaneous study of both 
        spin-dependent and spin-independent interactions of the
        DM particles with nuclei the chance for observing the
        DM signal is significantly increased.

	There are many different nuclear structure calculations 
	(including the case of non-zero momentum transfer) 
        for spin-dependent
	neutralino interaction with various nuclei, in particular with 
	helium $^3$He 
\cite{Vergados:1996hs},
	fluorine $^{19}$F
\cite{Vergados:2002bb,Divari:2000dc,Vergados:1996hs},
	sodium $^{23}$Na
\cite{Vergados:2002bb,Ressell:1997kx,Divari:2000dc,Vergados:1996hs},
	aluminium $^{27}$Al
\cite{Engel:1995gw},
	silicon $^{29}$Si
\cite{Vergados:2002bb,Ressell:1993qm,Divari:2000dc},
	$^{35}$Cl
\cite{Ressell:1993qm},
	potassium $^{39}$K
\cite{Engel:1995gw},
	germanium $^{73}$Ge 
\cite{Ressell:1993qm,Dimitrov:1995gc},
	niobium $^{93}$Nd
\cite{Engel:1992qb},
	iodide $^{127}$I
\cite{Ressell:1997kx},
	xenon $^{129}$Xe
\cite{Ressell:1997kx}, 
	$^{131}$Xe
\cite{Ressell:1997kx,Nikolaev:1993vw,Engel:1991wq}, 
	tellurium $^{123}$Te
\cite{Nikolaev:1993vw},
	tellurium $^{125}$Te
\cite{Ressell:1997kx},
	lead $^{208}$Pb
\cite{Kosmas:1997jm,Vergados:1996hs}.
	The zero-momentum transfer limits (mostly quenching) 
        were investigated for the target nuclei
	Cd, Cs, Ba and La in 
\cite{Pacheco:1989jz,Nikolaev:1993vw,Iachello:1991ut}.

\smallskip

        A dark matter event is an elastic scattering 
	of a relic neutralino $\chi$ (or $\tilde\chi$)\ 
	on the target nucleus $(A,Z)$, which results in a nuclear 
	recoil with energy $E_{\rm R}$ detected by a proper detector.
	The differential event rate in respect to the nuclear recoil 
	energy is a subject of experimental measurements.
	It depends on the distribution of
        the relic neutralinos in the solar vicinity $f(v)$ and
        the cross section of neutralino-nucleus elastic scattering
\cite{Jungman:1996df,Lewin:1996rx,Smith:1990kw,Bednyakov:1999yr,Bednyakov:1996yt,Bednyakov:1997ax,Bednyakov:1997jr,Bednyakov:1994qa}.
	The differential event rate per unit mass of 
	the target material takes the form
\begin{equation}
\label{Definitions.diff.rate}
	\frac{dR}{dE_{\rm R}} = N \frac{\rho_\chi}{m_\chi}
	\displaystyle
	\int^{v_{\rm max}}_{v_{\rm min}} dv f(v) v
	{\frac{d\sigma}{dq^2}} (v, q^2). 
\label{eq:1}
\end{equation}
	Here, $N={\cal N}/A$ is the a number density 
	of target nuclei. ${\cal N}$ and  $A$ stand for 
	the Avogadro number and the atomic mass in AMU, respectively.  
        The typical value of the nuclear recoil energy 
	$E_{\rm R} = q^2 /(2 M_A )$ is about $10^{-6} m_{\chi}$. 
        $M_A$ denotes the nuclear mass.

	The neutralino-nucleus elastic scattering cross section 
	for spin-non-zero ($J\neq 0$) nuclei is a sum of
        the coherent (spin-independent, or
	SI) and axial (spin-dependent, or SD) terms
\cite{Engel:1992bf,Engel:1991wq,Ressell:1993qm}: 
\begin{eqnarray}
\nonumber
{\frac{d\sigma^{A}}{dq^2}}(v,q^2) 
	&=& \frac{\sum{|{\cal M}|^2}}{\pi\, v^2 (2J+1)} 
         =  \frac{S^A_{\rm SD} (q^2)}{v^2 (2J+1)} 
           +\frac{S^A_{\rm SI} (q^2)}{v^2 (2J+1)} \\
\label{Definitions.cross.section}
        &=& \frac{\sigma^A_{\rm SD}(0)}{4\mu_A^2 v^2}F^2_{\rm SD}(q^2)
           +\frac{\sigma^A_{\rm SI}(0)}{4\mu_A^2 v^2}F^2_{\rm SI}(q^2).
\label{eq:2}
\end{eqnarray} 
	The normalized-to-unity ($F^2_{\rm SD,SI}(0) = 1$)
	nonzero-momentum-transfer nuclear form-factors
\begin{equation}
\label{Definitions.form.factors}
F^2_{\rm SD,SI}(q^2) = \frac{S^{A}_{\rm SD,SI}(q^2)}{S^{A}_{\rm SD,SI}(0)} 
\label{eq:3}
\end{equation}
	can be expressed through the nuclear structure functions as follows
\cite{Engel:1992bf,Engel:1991wq,Ressell:1993qm}: 
\begin{eqnarray}
\label{Definitions.scalar.structure.function}
S^{A}_{\rm SI}(q) 
	&=& 
	\sum_{L\, {\rm even}} 
        \vert\langle J \vert\vert {\cal C}_L(q) \vert\vert J \rangle \vert^2 
	\simeq  
	\vert\langle J \vert\vert {\cal C}_0(q) \vert\vert J \rangle \vert^2 ,
\nonumber \\ 
S^A_{\rm SD}(q) 
	&=& 
	\sum_{L\, {\rm odd}} \big( 
	\vert\langle N \vert\vert {\cal T}^{el5}_L(q) 
	\vert\vert N \rangle\vert^2 + \vert\langle N \vert\vert 
	{\cal L}^5_L (q) \vert\vert N \rangle\vert^2\big). 
\label{Definitions.spin.structure.function}
\label{eq:4}
\end{eqnarray} 
        Here, the double vertical lines denote  the reduced
	matrix element.
	The explicit form of the transverse electric ${\cal T}^{el5}(q)$ 
	and longitudinal ${\cal L}^5(q)$ multipole projections of the
	axial vector current operator, scalar function ${\cal C}_L(q)$ 
	and $S^{A}_{\rm SI,\, SD}(q)$ at zero momentum transfer   
	can be found in 
Appendix A.	
	For $q=0$  the nuclear SD and SI cross sections  
(in (\ref{Definitions.cross.section})) take the forms  
\begin{eqnarray}
\sigma^A_{\rm SI}(0) 
	&=& \frac{4\mu_A^2 \ S^{}_{\rm SI}(0)}{(2J+1)}\! =\!
	     \frac{\mu_A^2}{\mu^2_p}A^2 \sigma^{p}_{{\rm SI}}(0), \\ 
\sigma^A_{\rm SD}(0)
	&=&  \frac{4\mu_A^2 S^{}_{\rm SD}(0)}{(2J+1)}\! =\!
	     \frac{4\mu_A^2}{\pi}\frac{(J+1)}{J}
             \left\{a_p\langle {\bf S}^A_p\rangle 
                  + a_n\langle {\bf S}^A_n\rangle\right\}^2\\
      &=&
	\frac{\mu_A^2}{\mu^2_{p,n}}\frac{(J+1)}{3\, J}
\left\{ \sqrt{\sigma^{p}_{{\rm SD}}(0)}\langle{\bf S}^A_p\rangle 
       +{\rm sign}(a_p a_n)
	\sqrt{\sigma^{n}_{{\rm SD}}(0)}\langle{\bf S}^A_n\rangle\right\}^2.
\label{eq:5}
\end{eqnarray}
	Here, $\displaystyle \mu_A = \frac{m_\chi M_A}{m_\chi+ M_A}$
	is the reduced mass of the neutralino and the nucleus 
	and it is assumed that $\mu^2_{n}=\mu^2_{p}$.
	The dependence on effective neutralino-quark couplings 
	${\cal C}_{q}$ and ${\cal A}_{q}$
	in the underlying (SUSY) theory 
(see Appendix A) 	
\begin{equation}
{\cal  L}_{eff} = \sum_{q}^{}\left( 
	{\cal A}_{q}\cdot
      \bar\chi\gamma_\mu\gamma_5\chi\cdot
                \bar q\gamma^\mu\gamma_5 q + 
	{\cal C}_{q}\cdot\bar\chi\chi\cdot\bar q q
	\right)
      \ + ... 
\label{eq:6}
\end{equation}
	and on the spin ($\Delta^{(p,n)}_q$)
	and the mass ($f^{(p,n)}_q$) structure of nucleons 
	enter into these formulas via the zero-momentum-transfer 
	proton and neutron SI and SD cross sections: 
\begin{eqnarray}
\sigma^{p}_{{\rm SI}}(0) 
	= 4 \frac{\mu_p^2}{\pi}c_{0}^2,
&\qquad&
	c^{p,n}_0 = \sum_q {\cal C}_{q} f^{(p,n)}_q; \\
\sigma^{p,n}_{{\rm SD}}(0)  
 	=  12 \frac{\mu_{p,n}^2}{\pi}{a}^2_{p,n} 
&\qquad&
	a_p =\sum_q {\cal A}_{q} \Delta^{(p)}_q, \quad 
	a_n =\sum_q {\cal A}_{q} \Delta^{(n)}_q.
\label{eq:7}
\end{eqnarray}
	The factors $\Delta_{q}^{(p,n)}$, which parameterize the quark 
	spin content of the nucleon, are defined as
	$ \displaystyle 2 \Delta_q^{(n,p)} s^\mu 
	  \equiv 
          \langle p,s| \bar{\psi}_q\gamma^\mu \gamma_5 \psi_q    
          |p,s \rangle_{(p,n)}$.
	The total nuclear spin (proton, neutron) operator 
	is given by
\begin{equation}
 {\bf S}_{p,n} = \sum_i^A {\bf s}_{p,n} ({i}),
\label{eq:8}
\end{equation}
	where the index $i$ runs over all nucleons. 

	The expectation values of the spin and angular  
        operators are evaluated, as a rule, in their $z$-projection 
	by assuming the state with the  maximal value of 
        the angular momentum projection $M_J=J$:
\begin{equation}
\langle {\bf S} \rangle \equiv \langle N \vert {\bf S} \vert N \rangle
\equiv  \langle J,M_J = J \vert S_z \vert J,M_J = J \rangle.
\label{eq:9}
\end{equation}
	Thus $ \langle {\bf S}_{p(n)} \rangle $ is the total spin of protons 
	(neutrons) averaged over all nucleons of the nucleus $(A,Z)$.

        The mean velocity $\langle v \rangle$ of 
	the relic neutralinos of our Galaxy
	is about  $ 300~{\rm km/s} = 10^{-3} c$.
	Assuming 
        $q_{\rm max}R \ll 1$, where $R$ is the nuclear radius 
	and $q_{\rm max} = 2 \mu_A v$ is the maximum of the momentum 
	transfer in the process of the $\chi A$ scattering, the 
        spin-dependent matrix element  takes a simple form
({\em zero momentum transfer limit})\
\cite{Engel:1995gw,Ressell:1997kx}:
\begin{equation}
\label{Definitions.matrix.element}
 {\cal M} = C \langle N\vert a_p {\bf S}_p + a_n {\bf S}_n
 	\vert N \rangle \cdot {\bf s}_{\chi}
 	  = C \Lambda \langle N\vert {\bf J}
	 \vert N \rangle \cdot {\bf s}_{\chi}.
\label{eq:10}
\end{equation}
	Here, ${\bf s}_{\chi}$ denotes the spin of the neutralino, and 
\begin{equation}
 \Lambda = {{\langle N\vert a_p {\bf S}_p + a_n {\bf S}_n
\vert N \rangle}\over{\langle N\vert {\bf J}
\vert N \rangle}} =
{{\langle N\vert ( a_p {\bf S}_p + a_n {\bf S}_n ) \cdot {\bf J}
\vert N \rangle}\over{ J(J+1)
}}. 
\label{eq:11}
\end{equation}
        Note a coupling  of the spin of $\chi$ to the spin carried
	by the protons and the neutrons.  The uncertainties arising from 
        the electroweak and QCD scale physics are incorporated 
        in the factors $a_p$ and $a_n$. 
	The normalization factor $C$ involves the coupling
	constants, the  masses of the exchanged bosons and 
        the mixing parameters relevant to the 
        lightest supersymmetric particle (LSP), i.e., it is not
        related to the associated nuclear matrix elements
\cite{Griest:1988ma}.  
        The above conclusions concerning the spin-dependent part of the 
        neutralino-nucleus scattering amplitude are also valid for
        the amplitude of any Majorana WIMP-nucleus scattering process.
	In the limit of zero momentum transfer $q=0$ 
	the spin structure function in Eq. 
(\ref{Definitions.spin.structure.function}) reduces to the form
\begin{equation}
S^A_{\rm SD}(0) = 
	{2 J + 1\over{\pi}} \Lambda^2 J(J + 1). 
\label{eq:12}
\end{equation}
	The nuclear matrix element ${\cal M}$ in 
Eq.~(\ref{Definitions.matrix.element}) is often related to the
	matrix element of the nuclear magnetic moment, which
        also consists from the  matrix elements of the total proton and
        neutron spin operators:
\begin{equation}
\label{Definitions.magnetic.moment}
\mu =  \langle N\vert g_n^s {\bf S}_n + g_n^l {\bf L}_n +
	g_p^s {\bf S}_p + g_p^l {\bf L}_p\vert N \rangle.
\label{eq:13}
\end{equation}
	The {\it free particle} $g$-factors (gyromagnetic ratios) 
	are given (in nuclear magnetons) by:
\begin{equation}
	g_n^s = -3.826, \qquad 
	g_n^l = 0, 	\qquad
	g_p^s = 5.586,	\qquad
	g_p^l = 1. 
\label{eq:14}
\end{equation}
        The nuclear magnetic moment $\mu$ is often used as a benchmark 
        for the accuracy of 
	the calculation of ${\bf S}_p \,$ and ${\bf S}_n \,$ 
\cite{Ressell:1997kx,Ressell:1993qm}.  

	If the neutralino mass
        $m_{\chi}$ is larger than few tens of GeV, 
	the value of the product $qR$ is no longer negligible and 
	the so-called {\em  finite momentum transfer limit}\
	has to be considered in the case of the 
        neutralino scattering on medium-heavy and heavy nuclei.  
	The corresponding formalism is a generalization of
        that used for the description
	of weak and electromagnetic semi-leptonic interactions
	in nuclei. 
 	We shall follow the conventions of 
\cite{Ressell:1997kx,Ressell:1993qm}.
	There is an advantage to use the isospin 
	instead of the proton-neutron  representation when discussing 
	$\chi A$ scattering at finite momentum transfer.  
	By rewriting the isoscalar and isovector coupling constants as 
        $a_0 = a_n + a_p$ and $a_1 = a_p - a_n$, respectively,
        the spin-dependent cross section $S^A_{\rm SD}(q)$ decouples into
	the isoscalar term $S_{00}$, the isovector term $S_{11}$ 
	and the interference term $S_{01}$ as follows:
\begin{equation}
S^A_{\rm SD}(q) = a_0^2 S_{00}(q) + a_1^2 S_{11}(q) + a_0 a_1 S_{01}(q).
\label{eq:15}
\end{equation}
	The structure functions in 
Eq. (\ref{eq:15}) consist of
	the expectation values of operators 
	$j_L(qr) [ Y_L \sigma]^{L \pm 1}$ ($L$ even), 
	which depend on spin and spatial coordinates.
	Using the decomposition of $S^A_{\rm SD}(q)$ in (\ref{eq:15})
        one can obtain structure functions for a {$\chi$}
	of arbitrary composition.

\smallskip
	The cross section of neutralino-nucleus scattering
        at zero momentum transfer exhibits a strong 
	dependence on the details of the  nuclear ground state
\cite{Divari:2000dc}. The goal of this review is to 
        collect the results of different calculations
        and discuss their spread and relevance.

\section{The zero momentum transfer limit}
	Only nuclei with odd number of either protons or neutrons
	possess nonzero total nuclear spin. At first,          
	the independent single-particle shell model ({\bf ISPSM})
        was employed  by Goodman and Witten 
\cite{Goodman:1985dc} and later by others  
\cite{Drukier:1986tm,Ellis:1988sh,Smith:1990kw}
        to estimate the spin content of the nucleus
	for the detection of dark matter. 
	This model utilizes the shell structure of the nucleus,
        in particular the fact that if  certain magic numbers of nucleons
	occur in the nucleus, it exhibits 
	remarkable stability properties, and the ground state expectation
	values of the total spin $J$ and the parity of the nucleus can be 
	described by those of the extra nucleon.

	For nuclei whose angular momentum $J$ is given
	by a single neutron (proton) with spin $s$ and the
	orbital momentum $L$, 
${\bf J}={\bf L}+{\bf s}$, 
	(the even number of nucleons which remain form pairs 
         with the opposite angular momentum projection and zero spin, 
         i.e., they do not contribute to ${\bf J}$, we have 
\cite{Ellis:1988sh}:
\begin{equation}
\label{eq:16}
\langle{\bf S}^A_{n (p)}\rangle = { J (J +1) - L(L +1) +{3 \over 4} \over 2J +2},
~~\langle{\bf S}^A_{p (n)}\rangle=0.
\end{equation}
        In the ISPSM the entire angular momentum $J$ and the parity 
        of the odd-odd nucleus (A,Z) are identified with a single proton (Z-odd)  
        or neutron (Z-even) state.  Then the spin matrix elements are given
        by Eq. (\ref{eq:16}).

	The ISPSM offers only a rough estimate of the spin matrix elements.
	From 
Tables~\ref{Nuclear.spin.main.table.1-13}--%
       \ref{Nuclear.spin.main.table.209} it follows that 
	the ISPSM predictions significantly overestimate  
	the values obtained in realistic calculations.
	The ISPSM results are qualitatively good only for light nuclei with 
	the single nucleon  outside the closed shell (e.g., 
	$^{17}$O); however, they become increasingly poor for 
	heavier isotopes, especially for those with many particles
	outside the closed shells. The realistic calculations 
        take into account the complex structure 
        of the nuclear wave functions, the fact
	that the contributions to spin matrix elements both 
	from paired nucleons and unpaired ones cannot be
        neglected and the phenomenon that the free nucleon structure 
        coefficients are renormalized when nuclear medium effects are 
	relevant. 
	These effects are known to play an important role
	in the calculation of the 
	matrix elements of the magnetic dipole moment too 
(see, for example, \cite{Iachello:1991ut}). 

\smallskip
	There are several elaborated nuclear structure approaches 
        which lead to more accurate predictions  of spin        
	matrix elements associated with the dark matter detection     
        on nuclei in comparison with the ISPSM.
         A full list of these models, to our knowledge,
        includes the Odd Group Model ({\bf OGM}) 
\cite{Engel:1989ix} and the extended OGM ({\bf EOGM})
\cite{Engel:1989ix,Engel:1992bf} of Engel and Vogel, 
        the Interacting Boson Fermion Model ({\bf IBFM}) of
	Iachello, Krauss, and Maino 	 
\cite{Iachello:1991ut},
	the theory of Finite Fermi Systems ({\bf TFFS}) of 
	Nikolaev and Klapdor-Kleingrothaus
\cite{Nikolaev:1993dd}, the
	Quasi Tamm-Dancoff Approximation ({\bf QTDA}) of Engel
\cite{Engel:1991wq}, the nuclear shell model ({\bf SM})  applied  
        by Pacheco and Strottman 
\cite{Pacheco:1989jz},
	Engel, Pittel, Ormand and Vogel 
\cite{Engel:1992qb}, 
	Engel, Ressell, Towner and Ormand
\cite{Engel:1995gw}, 	
	Ressell et al.
\cite{Ressell:1993qm}, 
	Ressell and Dean
\cite{Ressell:1997kx};
	and by Kosmas, Vergados et al.
\cite{Vergados:1996hs,Kosmas:1997jm,Divari:2000dc} to different
        nuclear systems, 
	the so-called ``{\bf hybrid}'' model of Dimitrov, Engel and Pittel 
\cite{Dimitrov:1995gc}
	and the perturbation theory ({\bf PT}) based on calculations 
	of Engel et al.
\cite{Engel:1995gw}.
	 
        The ISPSM predictions are fairly accurate for near-closed-shell-nuclei,
	but further away they tend to overestimate the spin
	contribution to the magnetic moment. 
	In an open-shell nucleus, the last odd particle polarizes the
	other nucleons in the direction opposite to its own spin, 
	which results in a spin-quenching effect entirely absent in 
	the single-particle picture.  
	Denying the idea about importance of only the last odd nucleon
	Engel and Vogel arrived at the ``odd-group'' model
\cite{Engel:1989ix} by assuming   
	that the nuclear spin is carried by the ``odd''
	unpaired group of protons or neutrons and only one of either 
	$\langle{\bf S}^A_n\rangle$ or $\langle{\bf S}^A_p\rangle$ is non-zero.
	The odd-group spin matrix elements are expressed with 
	the measured nuclear magnetic moment $\mu$ as
\begin{equation}\label{Nuclear.odd-group}
\langle{\bf S}^A_p\rangle 
	= {\mu - g^l_{p} J \over g^s_{p} - g^l_{p}} 
         = {\mu - J \over 4.586}, \qquad\qquad
\langle{\bf S}^A_n\rangle 
	= {\mu - g^l_{n} J \over g^s_{n} - g^l_{n}} 
         = {- \mu \over 3.826},
\end{equation}
	where $g$ denote gyromagnetic factors of the free nucleon 
        [see (\ref{eq:14})].  
        The OGM has been found successful, e.g., in the case of 
         $^{29}$Si with $J={1\over2}$ and
	unpaired neutrons. 
	The experimental value of the nuclear magnetic moment 
	$\mu=-0.555$ implies $\langle{\bf S}^{29}_p\rangle\approx 0$, and 
	$\langle{\bf S}^{29}_n\rangle \approx 0.15$, which is
\cite{Jungman:1996df}
	in good agreement with the shell-model 
	calculation of Ressell et al.
\cite{Ressell:1993qm}. 
	The results of OGM calculations are collected in 
Tables~\ref{Nuclear.spin.main.table.1-13}--%
       \ref{Nuclear.spin.main.table.209}.
	We note that for $^{73}$Ge with a complex nuclear structure
	the odd-group model prediction disagrees with the realistic 
        calculation of
\cite{Ressell:1993qm,Dimitrov:1995gc}.

	The odd-group model is a significant improvement in comparison
        with the ISPSM. The weak points of this approach are the facts
	that the roles of small but not vanishing angular momenta of the even
	system and of the meson-exchange currents, which can
	renormalize the $g$ factors in 
Eq.~(\ref{Nuclear.odd-group}), are ignored. 
        Engel and Vogel improved the OGM 
\cite{Engel:1989ix}
	by using additional information about $\beta$-decay $ft$ values
	and measured magnetic momenta
	of ``mirror pairs'' for nuclear systems with ($A<50$).  
	For these nuclei they proposed to use two relations
\cite{Engel:1989ix}:
\begin{eqnarray}
(\frac{g_A}{g_V})^2
\left( \langle{\bf S}_{\rm odd}\rangle 
   - \langle{\bf S}_{\rm even}\rangle\right)^2 
   &=& \left(\frac{6170}{ft}-1\right)\frac{J}{J+1}, \nonumber\\
\mu^{}_{\rm IS} &=& J+ 0.76\left( 
       \langle{\bf S}_{\rm odd}\rangle 
      +\langle{\bf S}_{\rm even}\rangle\right) +\mu_x.
\end{eqnarray} 
	Here, $g_A$ and $g_V$ are the axial vector and 
	vector coupling constants, respectively,
        $\mu^{}_x$ is a small correction induced by heavy meson exchange
	and $\mu^{}_{\rm IS}$ is a sum of two mirror magnetic moments 
	(isoscalar moment).
	For free nucleons we have $g_A=1.25$ and $g_V=1.0$. 
	However, 
        in the nuclear matter due to the effect of renormalization
        the value $g_A/g_V=1.00\pm 0.02$ is often considered.
	For light nuclei the spin matrix elements
$\langle{\bf S}^{A}_p\rangle$ 
	and 
$\langle{\bf S}^{A}_n\rangle$ evaluated 
	within the extended odd group model (EOGM) 
\cite{Engel:1989ix} 
        are listed in 
Tables~\ref{Nuclear.spin.main.table.1-13}--%
      \ref{Nuclear.spin.main.table.29-41}. 
	We see that there is quite good agreement between the EOGM 
(with $g_A/g_V=1.00$) and
	the more sophisticated shell-model calculations 
	for light odd-even isotopes performed by Pacheco and Strottman
\cite{Pacheco:1989jz}.
	Their calculations for $A<16$ nuclei assumed the 
	Cohen-Kurath interaction
\cite{Cohen:1965qa} 
	and a complete basis within the $p$-shell model space.
	For $A>16$ the Reid interaction was considered
	and the basis consisted of all allowed states within the 
	1$s$-0$d$ shell-model space.
	The results of 
\cite{Pacheco:1989jz} are given in 
Tables~\ref{Nuclear.spin.main.table.1-13}--%
      \ref{Nuclear.spin.main.table.29-41}.	
	For heavier mirror nuclei with A close to 50 
	the shell-model calculations
	are difficult  due to a large amount of configurations 
        which have to be taken into account. 

	The light nucleus $^{27}$Al is one of the active ingredients 
	of a very high-resolution 
	and low-threshold 
	sapphire-crystal (Al$_2$O$_3$)-based detector for
        dark matter search.
	Engel, Ressell, Towner and Ormand
\cite{Engel:1995gw} performed 
        calculation of proton and neutron spin expectation values
	for this isotope with the help of 
        the Lanczos m-scheme shell-model code CRUNCHER
\cite{Resler:1988aa}. 
	The nucleus $^{27}\!$Al lies in the middle of the $sd$ shell and the 
	m-scheme basis for $^{27}\!$Al contains 80115 Slater determinants.  
	Good agreement between the calculated and measured 
        spectroscopy of excited states was achieved for this nucleus.  
        In addition, the experimental  value of the magnetic moment 
        $\mu_{\rm exp} = 3.6415~\mu_N$ 
        ($\mu_N = e\hbar/2m_p$ is the nuclear magneton)
        was reproduced in calculation well. 
	The theoretical value $\mu = 3.584~\mu_N$
        was obtained with the help of the free-particle $g$-factors
\cite{Engel:1995gw}. 
        We recall that calculation
        of the magnetic moment requires evaluation of the same spin
        matrix elements needed to determine      
        the WIMP structure functions at $q^2 = 0$.   
        The corresponding values of  
	$\langle{\bf S}^{27}_p\rangle$ and $\langle{\bf S}^{27}_n\rangle$ 
        are given in         
Table~\ref{Nuclear.spin.main.table.19-29}. 
        We note that the authors calculated structure functions $S(q)$ 
	of $^{27\!}$Al at $q\ne 0$  as well 
\cite{Engel:1995gw}. 	

	In the case of $^{39}$K the shell-model diagonalization needed for
        the calculation of the nuclear spin matrix
        elements requires severe truncations to the active model space. 
        The problem is that $^{39}$K is so near the boundary between the 
	{\it sd} and {\it pf} shells and
	excitations of particles into higher shells can have significant
	effects that are often not well simulated by effective operators.
        Thus, for this nucleus Engel, Ressell, Towner and Ormand 
\cite{Engel:1995gw} 
        used  an alternative scheme based on perturbation theory ({\bf PT}) 
	for the evaluation of spin matrix elements. 	
	It was successfully implemented in calculations of
	several spin-dependent observables in closed-shell-plus 
	(or minus)-one nuclei
\cite{Towner:1987aa}.  
	The details of the method and the calculations can be fobbed in Ref. 
\cite{Engel:1995gw}. The authors considered two
	different residual interactions. 
	One (denoted as $I$ in 
Table~\ref{Nuclear.spin.main.table.29-41}) 
	is related to the one-boson-exchange potential of the Bonn type,
	but it is limited only to four or five important meson exchanges.  
	The resulting interaction has a weak tensor-force component 
	typical of Bonn potentials.  
	The other (denoted as $II$ in 
Table~\ref{Nuclear.spin.main.table.29-41}) is represented by 
	full G-matrix elements of the Paris potential 
	parameterized in terms of sums over Yukawa functions 
	of various ranges and strengths. 
	Interaction $II$\ exhibits a strong tensor force.
        The quality of the wave functions obtained was judged in terms  
	of magnetic moments and Gamow-Teller matrix elements, 
	including meson-exchange currents, isobar
	currents, and other relativistic effects. 
        The results were presented for both isoscalar 
	and isovector magnetic moments and their sum, i.e., for
	the magnetic moment of $^{39}$K, whose value is given in  
Table~\ref{Nuclear.spin.main.table.29-41}.
        The magnetic moments calculated with the help of both interactions 
        differ only slightly from each other and showed good agreement
        with corresponding  experimental values.  
        The same nuclear wave functions of $^{39}$K were also used
	for the calculation of $\langle{\bf S}^{39}_p\rangle$ and 
	$\langle{\bf S}^{39}_n\rangle$ 
(Table~\ref{Nuclear.spin.main.table.29-41}) and the structure function $S(q)$
\cite{Engel:1995gw}. 
        It is worthwhile to notice that for both types of interactions 
	the zero-momentum transfer spin matrix elements 
	coincide well with those obtained within the phenomenological EOGM
\cite{Engel:1989ix}.  

	For dark matter targets constructed of heavy nuclei,
        in particular, Ge, I, and Xe, the first elaborated calculation of
	spin-dependent matrix elements relevant to WIMP scattering
        was performed by Iachello, Krauss, and Maino within 
	the Interacting Boson Fermion Model ({\bf IBFM})
\cite{Iachello:1991ut}. 	 
        The applied IBFM wave functions were tested in a 
	comprehensive analysis of excitation energies, 
	electromagnetic transition rates and 
	intensities of transfer reactions
\cite{Iachello:1991ut}. 
	In this model the total spin operator has the form
	$ S = \sum s^\pi_p + \sum s^\nu_n + s^{}_{p,n} $,
	where $s^{\pi(\nu)}_{p(n)}$ are the paired proton (neutron)
	spins, and $s^{}_{p(n)}$ is the remaining unpaired 
	proton (neutron) spin.
	To estimate the matrix elements of the paired nucleons
	one should know the structure of Cooper pairs (bosons),
	which were incorporated in the model by fitting the matrix
	elements of the magnetic moments.
	The authors  end with the conclusion
        that the ISPSM predictions are generally within 
	15\% of their results for the nucleon spin matrix elements
\cite{Iachello:1991ut}. 
	If quenching of free-particle coefficients in the nuclear 
	environment is considered, the value of the spin matrix 
        elements is reduced by an additional
	factor not exceeding 60\%.
	The results obtained are listed in 
Tables~\ref{Nuclear.spin.main.table.71-95}--%
	\ref{Nuclear.spin.main.table.113-141}.
        We note that the spin-dependent WIMP-nucleus cross section 
        associated with the IBFM spin matrix elements 
        is always smaller than 
	the ISPSM prediction but not more than a factor of 5
\cite{Iachello:1991ut}.
	While the IBFM can incorporate the dominant collective effects, 
	it has some difficulty in including the spin polarization, 
	which plays a crucial role in axial vector scattering. 
	Unfortunately, this approach cannot be readily applied 
        to the case of  nonzero momentum transfer
\cite{Divari:2000dc}. 

        In 
\cite{Divari:2000dc}
     	Vergados with co-authors 
	investigated the spin-dependent elastic neutralino scattering 
	with light nuclei $^{19}$F, $^{23}$Na, and $^{29}$Si.
	The spin contribution to the differential cross section was
        obtained by the shell model calculations in the $sd$ shell 
        using the Wildenthal interaction,
	which was developed and tested over many years. 
	This interaction is known to reproduce accurately 
	many nuclear observables for $sd$ shell nuclei. 
	The Wildenthal two-body matrix elements as well 
	as the single-particle energies are determined by 
	fits to experimental data in nuclei from $A=17$ to $A=39$.
        The shell-model wave functions used by the authors were tested 
        in the calculation of the low-energy spectra and 
	ground state magnetic moment.
	Rather good agreement between theoretical results and experimental 
        data was achieved. This fact increases the confidence level of the 
        calculated spin matrix elements which are listed in 
Tables~\ref{Nuclear.spin.main.table.11-21}--%
       \ref{Nuclear.spin.main.table.29-41}.
        It is worth mentioning that these spin matrix elements 
        are in good agreement with those of previous calculations 
\cite{Ressell:1993qm}. 
        The authors of Ref. 
\cite{Divari:2000dc}
        found $^{19}$F to be the most favorable target 
        for dark matter	search via spin-dependent interaction of
	relatively light dark matter particles. 
        It is favored due to the fact that the corresponding 
        spin matrix element 
	is not quenched and that various isospin 
	channels add coherently. Further, it was demonstrated 
        that the effect of the nuclear structure  on the elastic 
	scattering cross section of LSP with light nuclei
        (including $q\neq 0$ behavior) 
	is well understood, both for coherent and spin modes
\cite{Divari:2000dc,Vergados:2002bb}. 

	The Theory of Finite Fermi Systems ({\bf TFFS}) was used
 	by Nikolaev and Klapdor-Kleingrothaus
\cite{Nikolaev:1993dd}
	to describe spin matrix elements in the
	nuclear medium and to evaluate quenching of zero-momentum 
	nuclear spin matrix elements of heavy nuclei 
	due to residual interactions.
	Contrary to the OGM and the IFBM  studies the TFFS calculation
        of spin matrix elements is not related with the 
	experimental value of the associated magnetic moment. 
	The TFFS proton and neutron spin averages 
	$\langle {\bf S}_{p(n)} \rangle$ are suppressed in comparison with 
	the corresponding ISPSM predictions and  in some cases
	they differ significantly from the OGM values too
\cite{Engel:1989ix}. 
        However, they agree well  (with exception of the
	case of $^{73}$Ge)  
        with the results obtained by Pacheco and Strottman
\cite{Pacheco:1989jz} in a completely different IFBM approach 
(Tables~\ref{Nuclear.spin.main.table.71-95}--%
	\ref{Nuclear.spin.main.table.113-141}).

        The momentum transfer dependence 
        of the structure function $S(q)$ associated with scattaring 
        of dark matter particles from $^{131}$Xe, a promising heavy target 
        for the dark matter search experiment, was investigated by Engel
\cite{Engel:1991wq}
        by using the configuration-mixing
	quasiparticle Tamm-Dancoff approximation ({\bf QTDA}). 
	In the zeroth order the ground state of $^{131}$Xe was represented
	as the $1d_{3/2}$ quasineutron excitation of 
	the even-even core $|0\rangle$ treated in the BCS approximation 
	(BCS-based model of the Fermi surface).
	In the case of odd-multipole operators 
(\ref{Definitions.spin.structure.function}) the one-quasiparticle 
	approximation corresponds to the ISPSM approach of 
\cite{Ellis:1988sh}. 
	In order to incorporate nuclear structure corrections 
	originating from the residual interaction 
	three-quasiparticle configurations of the form 
$[\nu^{\dag}_{d3/2} [\nu^{\dag}_k\nu^{\dag}_l]^K ]^{3/2} |0\rangle$ and 
$[\nu^{\dag}_{d3/2} [\pi^{\dag}_k\pi^{\dag}_l]^K ]^{3/2} |0\rangle$ 
	were admixed. 
	Here $\pi^{\dag}$ and $\nu^{\dag}$ represent the proton and 
	neutron quasiparticle creation operators, $K$ is an arbitrary
	intermediate angular momentum, and $k,l$ run over
	a valence space consisting of the $2s$, $1d$, $0g$ and $0h$ 
	harmonic oscillator levels
\cite{Engel:1991wq}. 
	Despite the fact that the amplitudes associated with the 
        admixed three-quasiparticle states are small (less than 5\%),
	these admixtures can lead to a substantial effect.
	The experimental value of the  magnetic moment of $^{131}$Xe, which
	is about $0.69\, \mu_N$, 
	was reproduced with an accuracy of 2\% in the QTDA
	(note that the ISPSM value is almost twice larger). 
	The same approximation scheme results in 
	$\langle{\bf S}^{131}_p\rangle = -0.041$ 
	and $\langle{\bf S}^{131}_n\rangle = -0.236$
(Table~\ref{Nuclear.spin.main.table.128-133}).
 
	The neutralino-nucleus cross section associated with the 
        spin-dependent interaction is determined by the 
	distribution of the spin in the nucleus. 
	This observable is difficult to describe accurately as
	the lowest order nucleons create pairs with zero spin.
	The spin-dependent scattering mostly takes place near the nuclear
	Fermi surface and it is affected by 
	the behavior of relatively few nucleons.
	The silicon nucleus is  a relatively light. 
	Thus, shell-model calculations
	or arguments based on  existing data of magnetic moments
        (OGM and EOGM) allow a reliable prediction of deviations 
        from the simple ISPSM picture. 
	A different type isotope is
	germanium with a complex nuclear structure.
        For this nucleus the reliable calculation
        of spin expectation values is rather difficult.
	Niobium isotope $^{93}$Nb is something in between
        the above two special cases. 
	It is a heavy nucleus, which can be represented by a basic
        shell-model space corresponding to three 
	protons in the $1p_{1/2}$ or $0g_{9/2}$ levels and 
	two neutrons in the $1d_{5/2}$ level
\cite{Engel:1992qb}.
        This model space was considered by Engel, Pittel, Ormand and Vogel  
\cite{Engel:1992qb} 
        in the calculation of the magnetic moment for this isotope.
        They found $\mu$ to be equal to $6.36\, \mu_N$, which exceeds
        the experimental value $\mu~=~6.17\,\mu_N$. 
	In order to obtain better agreement with the experiment
        the authors considered a ``large'' model space which included all 
        basis states in which one proton or one neutron is excited 
        from the small model space. 
	Then, they ended up with the magnetic moment equal to $5.88\,\mu_N$, a
	value that is smaller. 
	The explanation could be that
        meson-exchange currents renormalize orbital proton $g$-factors 
	upwards by about 10\%, increasing the $\mu$ without
	altering the values of the nuclear spin matrix elements.
	The discrepancy between the $\langle{\bf S}^{93}_{p,n}\rangle$
	results obtained  within the small and large model spaces 
(Table~\ref{Nuclear.spin.main.table.71-95})
	provides an indication of 
	uncertainty of the nuclear-structure calculation 
\cite{Engel:1992qb}. 
 
	Germanium isotopes 
	(especially large-spin ${}^{73}$Ge) are considered to be the most 
	promising material for the direct dark matter search experiment. 
	However, there are fundamental difficulties in describing, 
	e.g.,  the spin content
	of ${}^{73}$Ge due to its complicated collective structure.
	Several studies were devoted  to nuclear structure aspects 
        of spin-dependent scattering of neutralinos from ${}^{73}$Ge.
	Engel and Vogel (OGM)
\cite{Engel:1989ix} used measured magnetic moments to
	estimate the quenching of the nucleon spin in several heavy nuclei, 
	including germanium.  
	Iachello, Krauss and Maino
\cite{Iachello:1991ut} employed the IBFM, and 
	Nikolaev and Klapdor-Kleingrothaus
\cite{Nikolaev:1993dd}
	used the TFFS to calculate the same quantities. 
        There are two most comprehensive spin structure
	analyses for ${}^{73}$Ge.
	A large-basis shell model study was performed by 
Ressell et al.
\cite{Ressell:1993qm},
	who calculated the  full spin-dependent neutralino response 
	including $q$-dependence of form factors.
	An equally comprehensive calculation 
	was realized by Dimitrov, Engel and Pittel 
\cite{Dimitrov:1995gc}. The authors obtained
        significantly different results in comparison 
        with other studies and they argue 
	that their results are more reliable than the previous ones.

	By using a reasonable two-body interaction Hamiltonian 
        and appropriate large model spaces 
Ressell et al. 
\cite{Ressell:1993qm}
	calculated the ground state wave functions for
        $^{29}$Si and $^{73}$Ge. In particular,
	the universal $sd$ shell interaction of Wildenthal 
	was used for calculation of the wave functions for silicon.
	The nuclear wave functions obtained were tested 
        in the analysis of energy pattern of excited states 
        and magnetic moments.
	Once reasonable agreement among theoretical results and 
	experimental data was obtained, the ground-state 
	wave functions were used to calculate the 
	neutralino-nucleus nuclear matrix elements.
        In addition, finite momentum transfer matrix elements and
        cross sections for the spin-dependent elastic scattering  
	of neutralinos from $^{29}$Si and $^{73}$Ge
\cite{Ressell:1993qm} were evaluated.  
	The computations were performed by using the Lanczos method 
	m-scheme nuclear shell model (code CRUNCHER)
\cite{Resler:1988aa}.
	The m-scheme basis for $^{29}$Si 
	had a dimension of 80115 Slater determinants. 
	In the limit of zero momentum transfer the 
	scattering matrix elements for $^{29}$Si are 
	in general agreement with previous estimates
(Table~\ref{Nuclear.spin.main.table.29-41}).

	For the study of $^{73}$Ge Ressell et al.
\cite{Ressell:1993qm} chose the 
	Petrovich-McManus-Madsen-Atkinson interaction
\cite{Petrovich:1969aa}, 
	which is a reasonable approximation to a full G-matrix calculation. 
	This interaction proved to be both adequate and tractable 
	in shell model applications. 
	Two different model spaces were considered.
	The  ``small'' space was determined by an m-scheme basis 
	dimension of 24731 Slater determinants. 
	The ``large'' space allowed much more excitations 
	with an m-scheme basis dimension of 117137 Slater determinants. 
	Despite fairly large size of the bases, 
	rather severe truncations in the space were enacted. 
	The small space is the smallest one in which it is possible
	to obtain agreement with the experimental spectrum energy levels. 
	The dimension of the large basis  was limited by the  computer 
	time and the memory storing constraints
\cite{Ressell:1993qm}. 
	No phenomenological interaction has been developed for 
	Ge-like nuclei and fairly 
	severe truncations to the model space have to be 
	imposed to obtain manageable dimensions.
	Despite these obstacles, the ground state 
	wave function for $^{73}$Ge allowed good description of
        the low-lying excited states and  
	the ground state to ground state spectroscopic factor.
        The large model space wave function of $^{73}$Ge led
        to an improved description of the ground state expectation
        values, in particular of the value of the magnetic moment,
        in comparison with the ISPSM and IBFM estimates. 
	The calculated magnetic moment $\mu$ from 
\cite{Ressell:1993qm} exceeds the experiment value, but the authors stressed 
	that the same quenching of both $\mu$ 
	and the Gammov-Teller (GT) spin matrix elements was almost 
	universally required in shell model calculations of 
        all heavy nuclei. 
	Assuming the isovector spin quenching factor  
	to be $0.833$, agreement with the measured $\mu$ is obtained.
	In principle, it is not obvious that quenching is really
	needed in neutralino-$^{71}$Ge scattering but if so, 
	Ressell et al. believed that the correct answer might
	be in the range between the quenched and unquenched values. 
	It was found 
(Table~\ref{Nuclear.spin.main.table.71-95}) 
	that the zero-momentum-transfer spin-neutron 
	matrix element 	$\langle{\bf S}^{73}_{n}\rangle$ of
        $^{73}$Ge was a factor of 2 larger than the previous predictions
	(except, obviously, the ISPSM value). 
	Thus,
	even if quenching is assumed, the calculated scattering rate 
	is about twice as large as any of the estimates made before
\cite{Ressell:1993qm}.

	A different sophisticated approach for evaluation of the spin 
	structure of $^{73}$Ge was considered by Dimitrov, Engel and Pittel.
	It relies on the idea of mixing variationally determined
	Slater determinants, in which symmetries are broken 
	but restored either before or after variation.  
        This approach is described in detail in
\cite{Dimitrov:1994aa}.  
	In the calculation of
\cite{Dimitrov:1995gc} 
	the symmetries broken in the intrinsic states are those 
	associated with rotational invariance, parity, and axial shape.  
	The hybrid procedure used restores axial symmetry,
	parity invariance, and approximate rotational invariance prior to
	the variation of each intrinsic state.   
	Subsequently, before mixing the intrinsic states the
	rotational invariance is fully restored.
	The procedure allows fully triaxial Slater
	determinants at the expense of particle-number breaking.  
	The results of 
\cite{Dimitrov:1994aa} indicate that the trading of number 
	nonconservation for triaxiality is a good idea, 
	despite the apparent loss of pairing correlations
	traditionally associated with the former.  
	Pairing forces evidently induce effective triaxiality.  
	The numerical results
\cite{Dimitrov:1994aa} show that the approach is
	accurate and efficient for describing even-even systems 
	while also providing reliable reproduction 
	of the collective dynamics of odd-mass systems
\cite{Dimitrov:1995gc}.

	For ${}^{73}$Ge the calculations were performed  
        by assuming, both for protons and neutrons, a single-particle model 
	consisting of the full $0f,1p$ shell and  
	the $0g_{9/2}$ and $0g_{7/2}$ levels.
	The main idea was to include all of the single-particle orbits that
	could play an important role in reproducing 
	low-energy properties of the ${}^{73}$Ge 
\cite{Dimitrov:1995gc}.
	It is well-known that a crucial 
        ingredient in any realistic nuclear-structure 
	calculation is the appropriate form of the nuclear Hamiltonian.  
	The one- and two-body parts of the Hamiltonian 
	have to be compatible with each other as well as with the model space.
	This is difficult to achieve because 
	microscopic two-body interactions, derived for example from a
	G-matrix, include monopole pieces that are unable 
	to describe the movement of spherical single-particle 
	levels as one passes from the beginning to the end of a shell.
	A  proposed procedure for avoiding this problem 
	consists basically in removing all monopole components 
	from the two-body interaction and 
	shifting their effects to the single-particle energies.  
	This procedure was used by Dimitrov, Engel and Pittel 
\cite{Dimitrov:1995gc} --- their two--body
	force was a fit to Paris-potential G-matrix
	modified as just described above.
	The calculated ground-state magnetic dipole moment is in
	good agreement with the experimental value.  
	Ressell et al. 
\cite{Ressell:1993qm} 
	in their large space shell model
	calculation were able to reduce $\mu$ significantly to 
	$-1.24 \mu_N$ (without direct quenching)
	but could not account for the remaining difference.  
	On the contrary, the calculation of Dimitrov, Engel and Pittel, 
	despite the small number of intrinsic states, 
	contains the full quenching required by experiment
\cite{Dimitrov:1995gc}.
	By making a comparison with the results of Ressell et al. again 
\cite{Ressell:1993qm}, 
        significant disagreement is found for the neutron spin. 
	The calculated value  of Dimitrov, Engel and Pittel 
	is significantly smaller
(Table~\ref{Nuclear.spin.main.table.71-95}).  
	The large and negative neutron spin $g$-factor ($g^s_n=-3.826$) 
	is favored  by the correct $\mu$ value.  
	The differences in the spins, unlike those in the 
	orbital angular momenta, carry over into WIMP scattering 
	cross sections. 
	Thus, following Ressell et al. 
\cite{Ressell:1993qm}, 
	no significant increase 
	is expected in the neutralino-${}^{73}$Ge scattering rate.

	The advantage of Dimitrov, Engel and Pittel's approach 
	for calculation of neutralino cross sections 
	is that it correctly represents the spin structure, 
	requires neither quenching at $q=0$ nor arbitrary assumptions about 
	the form factor behavior at $q \neq 0$
\cite{Dimitrov:1995gc}.  
	The spin matrix elements depend in general rather sensitively 
	on the details of the nuclear structure.
	Since the matrix elements at $q=0$ are often quenched, 
	the momentum dependence of the matrix elements was 
	more important than it was naively expected. 
	As a matter of fact, one has to include a lot 
	of configurations to accommodate all multipoles, which 
	result in very large Hilbert spaces
	in complex nuclei like $^{29}$Si and $^{73}$Ge. 
	It will be therefore a very hard task to substantially improve
	the calculations of Ressell et al. 
\cite{Ressell:1993qm} and Dimitrov, Engel and Pittel
\cite{Dimitrov:1995gc} for these elements.

	For evaluations of the spin matrix elements in the heaviest 
	possible nuclei relevant to dark matter search 
	Kosmas and Vergados have chosen 
	$^{207}$Pb 
\cite{Vergados:1996hs,Kosmas:1997jm}. 
	Among the targets which were considered for 
	direct neutralino detection, 
	$^{207}$Pb stands out as an important candidate. 
	The spin matrix element of this nucleus has not been evaluated
	quite accurately, since one expected 
	that the neutralino spin interaction 
	is important only with light nuclei. 
	But the spin matrix element in the light systems is quenched. 
	On the other hand, the spin matrix element of $^{207}$Pb, 
	especially the isoscalar one, does not suffer unusually large 
	quenching, as is known from the study of the magnetic moment. 
	It is believed that $^{207}$Pb has a quite simple structure, 
	its ground state can be described as a $2p_{1/2}$ neutron hole 
	outside the doubly magic (closed-shell) nucleus $^{208}$Pb.
	Due to its low angular momentum, only two multipoles $L =0$ and $L =2$ 
	can contribute even at large momentum transfers. 
	One can thus view the information obtained from 
	this simple nucleus as complementary to that of 
	$^{73}$Ge, which has a very complex nuclear structure
\cite{Vergados:1996hs,Kosmas:1997jm}.
	In the $q=0$ limit Vergados and Kosmas gave
	the spin matrix element in the simple form 
$|{\bf J}|^2 = \Big|f^0_A \Omega_0(0) + f^1_A \Omega_1(0)\Big|^2$,
	and found that
	$\Omega_0(0) =-0.95659/\sqrt{3}$, and 
	$\Omega_1(0) = 0.83296/\sqrt{3}$
\cite{Vergados:1996hs,Kosmas:1997jm}.
	These values were recalculated in the form of spin variables 
	$ \langle {\bf S}_{p(n)} \rangle$ given in
Table~\ref{Nuclear.spin.main.table.209}. 

	Ressell and Dean 
\cite{Ressell:1997kx} have performed most accurate nuclear shell model 
	calculations of the neutralino-nucleus spin-dependent or axial
	cross section for several nuclei in the $A=127$ region,
	which are important for dark matter search. 
	Their set of structure functions $S(q)$
	is valid for all relevant values of the momentum transfer. 
	Conventional nuclear shell model of Wildenthal 
\cite{Brown:1988aa} quite accurately represents spin-dependent
	neutralino-nucleus matrix elements 
	when a reasonable nuclear Hamiltonian is used in a sufficiently
	large model space  
\cite{Ressell:1997kx}.
	Until recently, both of these ingredients have been absent for nuclei
	in the $3s2d1g_{7/2}1h_{11/2}$ shell.
	Ressell and Dean considered two residual nuclear interactions 
	based upon recently developed realistic nucleon-nucleon 
	Bonn A 
\cite{Hjorth-Jensen:1995ap} and Nijmegen II 
\cite{Stoks:1994wp} potentials.  
	These two nucleon-nucleon 
	potentials were used 
	in order to investigate the sensitivity of the results 
	to the particular nuclear Hamiltonian.

	The Bonn-A-based Hamiltonian has been derived for the model 
	space consisting of the
	$1g_{7/2}, \, 2d_{5/2}, \, 3s_{1/2}, \, 2d_{3/2},$ and
	$1h_{11/2}$ orbitals, 
	allowing one to include all relevant correlations.
	In order to get good agreement
	with observables for nuclei with $A \approx 130$, 
	the single-particle energies (SPEs) were adjusted.  
	The SPEs were varied 
	until reasonable agreement between calculation
	and experiment was found for the magnetic moment, 
	the low-lying excited state energy
	spectrum, and the quadrupole moment 
	of $^{127}$I.  
  	Once the SPEs are specified, 
	a reasonable Hamiltonian can be used for the nuclei 
	under investigation.

	To perform a full basis calculation of the
	$^{127}$I ground state properties in the space consisting of the
	$1g_{7/2}, \, 2d_{5/2}, \, 3s_{1/2}, \, 2d_{3/2},$ and
	$1h_{11/2}$ orbitals, one would need basis states consisting
	of roughly $1.3 \times 10^9$ Slater Determinants (SDs).
	Current calculations 
	can diagonalize matrices with basis dimensions in the range
	1--2$\times 10^7$ SDs.
	Therefore clearly severe truncations of the model space are needed
\cite{Ressell:1997kx}.  
	Fortunately, given the size of the model
	spaces that can be treated, a truncation scheme that includes
	the majority of relevant configurations can be devised.  
  	Finally (after relevant truncations, see details in 
\cite{Ressell:1997kx}) the m-scheme dimension of
	the $^{127}$I  model space is about 3 million SDs.
	The calculated observables agree well with experiment.
	These interactions do not seem to prefer
	excitation of more than one extra neutron pair to 
	the $1h_{11/2}$.  
	Most configurations have six neutrons in that orbital, 
	while eight are allowed.  
	Hence, this model space is more than adequate.
	It is this truncation scheme that was used 
	for the two Xenon isotopes considered ($A = 129$ and $131$). 

	In almost every instance, the results of 
\cite{Ressell:1997kx}
(Tables~\ref{Nuclear.spin.main.table.114-125}--%
	\ref{Nuclear.spin.main.table.128-133})
	show that the spin
	$\langle {\bf S}_i \rangle $ ($i = p,n$) carried by the unpaired
	nucleon is greater than that found in the other nuclear
	models (except for the ISPSM, where $\langle {\bf S}_i \rangle $
	is maximal).  
	Despite these larger values for $\langle {\bf S}_i \rangle $,
	these calculations have significant quenching of the magnetic moment
	and are in good agreement with experiment in all cases.
	The larger values of $\langle {\bf S}_i \rangle $ 
	are due to the fact that more excitations of the
	even group of the nuclei were allowed
\cite{Ressell:1997kx}.
	The differences in the response due to the two forces is clearly
	visible in 
Tables~\ref{Nuclear.spin.main.table.114-125}--%
       \ref{Nuclear.spin.main.table.128-133}.
	In all cases reasonable
	agreement between calculation and experiment for the magnetic
	moment (using free particle $g$-factors) is achieved.
	It is obvious 
	that the differences between the two calculations
	are non-trivial but they are quite a bit smaller
	than the differences coming from the use of alternate nuclear models.  
	This shows that the interaction is not the primary
	uncertainty in calculations of the 
	neutralino-nucleus spin cross sections
\cite{Ressell:1997kx}.

	The results obtained by Ressell and Dean give a factor 
	of $20$ increase in iodine's sensitivity to spin-dependent
	scattering over that previously assumed.  
	Due to the form factor suppression a sodium
	iodide detector's spin response is still
	dominated by $^{23}$Na but not to the extent previously thought.
	For the remainder of the nuclei considered 
Tables~\ref{Nuclear.spin.main.table.114-125}--%
       \ref{Nuclear.spin.main.table.128-133}  
	also reveal increased scattering sensitivity, though much more modest
\cite{Ressell:1997kx}.

	Before finishing this section we, following Ressell and Dean
\cite{Ressell:1997kx}, 
	discuss the quenching problem and some related uncertainties.
	As is already noted above, the comparison of the computed
	magnetic moment and its experimental value has been
	used as the primary indicator of the calculation's reliability.  
	This seems quite reasonable in the light of the 
	similarities between the matrix elements in 
Eqs.
(\ref{Definitions.matrix.element}) and 
(\ref{Definitions.magnetic.moment}).  
	This prescription is not free of several potential problems 
\cite{Ressell:1993qm,Engel:1995gw}.  
	Not only does $\mu$ depend
	upon the orbital angular momentum ${\bf L}_i$ but the spin
	angular momentum  ${\bf S}_i$ is subtly different.
	The neutralino-nucleus matrix element 
(\ref{Definitions.matrix.element}) results from the non-relativistic
	reduction of the axial-vector current.  
	Because of this, it is
	not strongly affected by meson exchange currents (MECs).  
	The magnetic moment's spin operators, ${\bf S}_i$, are a result of the
	non-relativistic reduction of the vector current.  
	They can be strongly affected by MECs 
\cite{Engel:1995gw}.
	The effects of MECs upon $\mu$ is typically lumped together with
	several other effects to give effective $g$-factors.
	Unfortunately, there is no hard and
	fast rule as to what effective $g$-factors are the best.  
	One usually chooses to remain with the free particle
	$g$-factors.  
	As an example of the potential uncertainties this
	ambiguity leads to, 
	the calculated magnetic moments for these
	nuclei based on a reasonable set of effective $g$-factors
	were also included in 
Tables~\ref{Nuclear.spin.main.table.114-125}--%
       \ref{Nuclear.spin.main.table.128-133}.  
	The ``quenched'' magnetic moments are the values
	in curled parentheses and the effective $g$-factors used
	are $g_n^s = -2.87, \, g_n^l = -0.1, \, g_p^s = 4.18$, and
	$g_p^l = 1.1$.  
	The tables show that these $g$-factors do little
	to improve the concordance 
	between calculation and experiment
\cite{Ressell:1997kx}.
	A related concern involves the quenching of the 
	(isovector) Gamow-Teller (GT) $g$-factor, $g_A$ 
\cite{Ressell:1993qm,Engel:1995gw}.  
	The spin term of the GT operator also comes from the 
	axial vector current and thus is closely related to 
	the spin operators in Eq.
(\ref{Definitions.magnetic.moment}).
	Its is well established that most nuclear model calculations of
	GT strength require a reduction of the order of 20\% in $g_A$ 
\cite{Brown:1988aa}. 
	Whether this quenching of $g_A$ should also be applied 
	to $a_1$ (the isovector neutralino-nucleon coupling constant) 
	is unknown.  
	Since there is no real guidance and magnetic moments 
	obtained by Ressell and Dean agree well with experiment, 
	it is very doubtful that any extra quenching 
	of the spin matrix elements (or equivalently the
	coupling constants $a_0$ and $a_1$) is desirable for these nuclei
	in the calculation of neutralino-nucleus scattering rates.  
	Nonetheless, it is useful to keep these potential
	uncertainties in mind when calculating scattering rates
\cite{Ressell:1993qm,Engel:1995gw,Ressell:1997kx}.

Tables~\ref{Nuclear.spin.main.table.1-13}--%
       \ref{Nuclear.spin.main.table.209} 
	contain the fullest possible list of
	calculations of the nuclear zero-momentum spin properties
	considered in the literature for detection of spin-coupled WIMPs.
	The tables are obtained on the basis of relevant tables for 
	$\langle{\bf S}^A_n\rangle$  and $\langle{\bf S}^A_p\rangle$
	from 
\cite{Engel:1989ix,Engel:1991wq,Engel:1992bf,Engel:1992qb,Engel:1995gw,%
Dimitrov:1995gc,Pacheco:1989jz,Iachello:1991ut,Nikolaev:1993vw,Jungman:1996df,%
Ressell:1993qm,Ressell:1997kx,Divari:2000dc}.
	The OGM results of Ellis and Flores given in 
\cite{Ellis:1988sh,Ellis:1991ef,Ellis:1993vh} in
	the form of $\lambda^2 J(J+1)$ were recalculated 
	into $\langle {\bf S}_i \rangle$ and checked. 

\section{Summary and Conclusions} 
       There is continuous theoretical and experimental 
       interest in existence of dark matter of the Universe. 
       The best motivated non-baryonic dark matter candidates 
       is the neutralino, the lightest supersymmetric particle. 
       The motivation for supersymmetry arises
       naturally in modern theories of particle physics. 
       In this work we discussed the spin-dependent interaction 
       of neutralinos with odd mass nuclei. 
       The nuclear structure plays an important role  in determining 
       the strength of the neutralino-nucleus cross section for this
       type of interaction. 
       In the limit of zero momentum transfer  
       the relevant physical quantities are 
       the proton and neutron spin averages $ \langle {\bf S}_{p(n)} \rangle $,
       which have to be evaluated within a proper nuclear model. 
       These values determine the event rate expected 
       in a direct dark matter search experiment  
       due to spin-dependent neutralino-nucleus interaction.
       In this work the calculation of spin-dependent matrix elements is
       reviewed. 
       To our knowledge, a complete list of calculated
       spin matrix elements is presented for nuclei throughout 
       the periodic table.
       We recall that only nuclei with an odd number of either protons or 
       neutrons can have non-zero spin. 

       As is manifested in this review, practically every known 
       nuclear model has been employed
       for evaluation of the  spin matrix elements. 
       The results show that spin matrix elements depend in general rather 
       sensitively on the details of the nuclear structure. 
       The phenomenological ISPSM (independent single-particle shell model)
       is fairly accurate only for light nuclei
       with near-closed shells, but in general it tends to overestimate 
       the spin matrix elements and is inadequate. 
       It was confirmed by the calculations within the OGM
       (odd group model) and the EOGM (extended OGM) utilizing 
       magnetic moments and mirror $\beta$ decays. 
       However, detailed shell model calculations found
       these phenomenological models to be inadequate.
       The odd group model and shell model treatments yielded 
       good agreement for light nuclei, but as the atomic mass increases, 
       there arouses a significant amount of configuration mixing
       not considered in the OGM. 
       Unfortunately, the shell model calculations are difficult for
       most medium-heavy and heavy isotopes
       because of the size of the matrices involved. 
       The situation is improving due to advances in computer 
       power and storage. 
       There is a hope to construct model spaces that contain most of the 
       nuclear configurations that are likely to dominate the 
       spin response of nuclei. 
       For open-shell medium-heavy and heavy nuclei the methods of choice
       for calculation of spin matrix elements are  
       the Interacting Boson-Fermion Model, the Theory of Finite Fermion
       Systems and the Quasi Tamm-Dancoff Approximation. 
       The most reliable values of $\langle {\bf S}_{p(n)} \rangle $  
       are considered to be those of the approach which 
       reproduces well the experimental value of the magnetic moment
       for a given isotope.
       The magnetic moment is extremely important, 
       as it is the observable most closely related to the 
       neutralino-nucleus scattering matrix element and has
       traditionally been used as a benchmark for 
       the calculation accuracy. 

       There is an additional complication arising 
       from the fact that the neutralino appears to be quite massive, 
       perhaps heavier than 100 GeV. 
       For such a heavy light supersymmetric particle
       and sufficiently heavy nuclei, the dependence of the nuclear matrix 
       elements on the momentum transfer cannot be ignored. 
       This affects the spin matrix elements. 
       The calculations of the structure
       functions in the finite momentum approximation
       and the level of accuracy of these calculations
       are beyond the scope of this review.

       This work was supported in part by 
       the VEGA Grant Agency of the Slovac Republic under contract
       No. 1/0249/03 and by the Russian Foundation for Basic Research
       (grant 02--02--04009).

\bigskip

\begin{table}[!ht] 
\caption{Zero momentum spin structure of light nuclei 
	($A<13$) in different models. 
	The measured magnetic moments used as input 
	are enclosed in parentheses.
\label{Nuclear.spin.main.table.1-13}}
\begin{center}
\begin{tabular}{lrrr}
\hline\hline 
$^{1}$H~($L_J=S_{1/2}$) & 
~~~~~~~~$\langle {\bf S}_p \rangle$ & 
~~~~~~~~$\langle {\bf S}_n \rangle$ &~~~~~~~~$\mu$ (in $\mu_N$) \\ \hline
ISPSM, Ellis--Flores~\cite{Ellis:1988sh,Ellis:1991ef}
	& $1/2$ & $0$ & $2.793$  \\ 
OGM, Ellis--Flores~\cite{Ellis:1988sh,Ellis:1991ef}
	& $0.5$ & $0$ & $(2.793)_{\rm exp}$ \\ 
\hline
\hline 
$^{3}$He~($L_J=S_{1/2}$) & 
$\langle {\bf S}_p \rangle$ & $\langle {\bf S}_n \rangle$ & 
~~~~~~~~$\mu$ (in $\mu_N$) \\ \hline
ISPSM, Ellis--Flores~\cite{Ellis:1988sh,Ellis:1991ef}
	& $0$ & $1/2$ & $-1.913$  \\ 
OGM, Engel--Vogel~\cite{Engel:1989ix} 
	& $0$ & $0.56$ & $(-2.128)_{\rm exp}$ \\ 
EOGM ($g_A/g_V=1$), Engel--Vogel~\cite{Engel:1989ix} 
	& $-0.081$ & $0.552$ & $(-2.128)_{\rm exp}$ \\ 
EOGM ($g_A/g_V=1.25$), Engel--Vogel~\cite{Engel:1989ix} 
	& $-0.021$ & $0.462$ & $(-2.128)_{\rm exp}$ \\ 
\hline
\hline
$^7$Li~($L_J=P_{3/2})$  &
$\langle {\bf S}_p \rangle$ & $\langle {\bf S}_n \rangle$ & 
~~~~~~~~$\mu$ (in $\mu_N$) \\ \hline
ISPSM, Ellis--Flores~\cite{Ellis:1988sh,Ellis:1991ef}
	& $1/2$ & $0$ & $3.793$  \\ 
OGM, Engel--Vogel~\cite{Engel:1989ix} 
	& $0.38$ & $0$ & $(3.256)_{\rm exp}$ \\ 
SM, Pacheco-Strottman~\cite{Pacheco:1989jz}
	& 0.497   & 0.004 \\
\hline
\hline
$^{9}$Be~($L_J=P_{3/2}$) & 
$\langle {\bf S}_p \rangle$ & $\langle {\bf S}_n \rangle$ & 
~~~~~~~~$\mu$ (in $\mu_N$) \\ \hline
ISPSM, Ellis--Flores~\cite{Ellis:1988sh,Ellis:1991ef}
	& $0$ & $1/2$ & $-1.913$  \\ 
OGM, Engel--Vogel~\cite{Engel:1989ix} 
	& $0$ & $0.31$ & $(-1.178)_{\rm exp}$ \\
SM, Pacheco-Strottman~\cite{Pacheco:1989jz}
	&0.007   &0.415 \\ 
\hline
\hline
$^{11}$B~($L_J=P_{3/2}$) & 
$\langle {\bf S}_p \rangle$ & $\langle {\bf S}_n \rangle$ & 
~~~~~~~~$\mu$ (in $\mu_N$) \\ \hline
ISPSM, Ellis--Flores~\cite{Ellis:1988sh,Ellis:1991ef}
	& $1/2$ & $0$ & $3.793$  \\ 
OGM, Engel--Vogel~\cite{Engel:1989ix} 
	& $0.264$ & $0$ & $(2.689)_{\rm exp}$ \\ 
EOGM ($g_A/g_V=1$), Engel--Vogel~\cite{Engel:1989ix} 
	& $0.292$ & $0.006$ & $(2.689)_{\rm exp}$ \\  
EOGM ($g_A/g_V=1.25$), Engel--Vogel~\cite{Engel:1989ix} 
	& $0.264$ & $0.034$ & $(2.689)_{\rm exp}$ \\  
SM, Pacheco-Strottman~\cite{Pacheco:1989jz}
  	& $0.292$ & $0.008$ \\
\hline
\hline
\end{tabular} \end{center}
\end{table} 

\begin{table}[!ht] 
\caption{Zero momentum spin structure of light nuclei 
	($11<A<21$) in different models. 
\label{Nuclear.spin.main.table.11-21}}
\begin{center}
\begin{tabular}{lrrr}
\hline
\hline
$^{13}$C~($L_J=P_{1/2}$) & 
~~~~~~~~$\langle {\bf S}_p \rangle$ & 
~~~~~~~~$\langle {\bf S}_n \rangle$ & ~~~~~~~~$\mu$ (in $\mu_N$) \\ \hline
ISPSM 
	& 0 & $-0.167$ & $0.638$  \\ 
OGM 
	& 0 & $-0.183$ & $(0.702)_{\rm exp}$ \\ 
EOGM ($g_A/g_V=1$), Engel--Vogel~\cite{Engel:1989ix} 
	& $-0.009$ & $-0.172$ & $(0.702)_{\rm exp}$ \\   
EOGM ($g_A/g_V=1.25$), Engel--Vogel~\cite{Engel:1989ix} 
	& $-0.026$ & $-0.155$ & $(0.702)_{\rm exp}$ \\ 
\hline
\hline 
$^{15}$N~($L_J=P_{1/2}$) & 
~~~~~~~~$\langle {\bf S}_p \rangle$ & 
~~~~~~~~$\langle {\bf S}_n \rangle$ & 
~~~~~~~~$\mu$ (in $\mu_N$) \\ \hline
ISPSM, Engel--Vogel~\cite{Engel:1989ix} 
	& $-0.167$ & $0$ & $-0.264$  \\ 
OGM, Engel--Vogel~\cite{Engel:1989ix} 
	& $-0.167$ & $0$ & $(-0.283)_{\rm exp}$ \\ 
EOGM ($g_A/g_V=1$), Engel--Vogel~\cite{Engel:1989ix} 
	& $-0.145$ & $0.037$ & $(-0.283)_{\rm exp}$ \\   
EOGM ($g_A/g_V=1.25$), Engel--Vogel~\cite{Engel:1989ix} 
	& $-0.127$ & $0.019$ & $(-0.283)_{\rm exp}$ \\ 
\hline
\hline 
$^{17}$O~($L_J=D_{5/2}$) & 
$\langle {\bf S}_p \rangle$ & $\langle {\bf S}_n \rangle$ & 
~~~~~~~~$\mu$ (in $\mu_N$) \\ \hline
ISPSM, Ellis--Flores~\cite{Ellis:1988sh,Ellis:1991ef}
	& 0 		& $1/2$ 	& $-1.913$  \\ 
OGM, Engel--Vogel~\cite{Engel:1989ix} 
	& 0 		& $0.49$ 	& $(-1.894)_{\rm exp}$ \\ 
EOGM ($g_A/g_V=1$), Engel--Vogel~\cite{Engel:1989ix} 
	& $-0.036$ 	& $0.508$ 	& $(-1.894)_{\rm exp}$ \\ 
EOGM ($g_A/g_V=1.25$), Engel--Vogel~\cite{Engel:1989ix} 
	& $0.019$ 	& $0.453$ 	& $(-1.894)_{\rm exp}$ \\ 
SM, Pacheco-Strottman~\cite{Pacheco:1989jz}
	&0	    &0.5   \\
\hline
\hline 
$^{19}$F~($L_J=S_{1/2}$) & 
~~~~~~~~$\langle {\bf S}_p \rangle$ & 
~~~~~~~~$\langle {\bf S}_n \rangle$ & ~~~~~~~~$\mu$ (in $\mu_N$) \\ \hline
ISPSM, Ellis--Flores~\cite{Ellis:1988sh,Ellis:1991ef}
	& $1/2$ 	& $0$ 	& $2.793$  \\ 
OGM, Engel--Vogel~\cite{Engel:1989ix} 
	& $0.46$ 	& $0$ 	& $(2.629)_{\rm exp}$ \\ 
EOGM ($g_A/g_V=1$), Engel--Vogel~\cite{Engel:1989ix} 
	& $0.415$ 	& $-0.047$ & $(2.629)_{\rm exp}$ \\  
EOGM ($g_A/g_V=1.25$), Engel--Vogel~\cite{Engel:1989ix} 
	& $0.368$ 	& $-0.001$ & $(2.629)_{\rm exp}$ \\ 
SM, Pacheco-Strottman~\cite{Pacheco:1989jz}
	& $0.441$   	&$-0.109$  \\
SM, Divari et al.~\cite{Divari:2000dc} 
	& $0.4751$ 	& $-0.0087$ 
	& $2.91$  \\  
\hline
\hline 
\end{tabular} \end{center}
\end{table} 

\begin{table}[!ht] 
\caption{Zero momentum spin structure of light nuclei 
	($19<A<29$) in different models. 
\label{Nuclear.spin.main.table.19-29}}
\begin{center}
\begin{tabular}{lrrr}\hline\hline
$^{21}$Ne~($L_J=P_{3/2}$)& 
~~~~~~~~$\langle {\bf S}_p \rangle$ & 
~~~~~~~~$\langle {\bf S}_n \rangle$ & ~~~~~~~~$\mu$ (in $\mu_N$) \\ \hline
ISPSM 
	& 0 	& $1/2$ 	& $-1.913$  \\ 
OGM 
	& 0 	& $0.173$ 	& $(-0.662)_{\rm exp}$ \\ 
EOGM ($g_A/g_V=1$), Engel--Vogel~\cite{Engel:1989ix} 
	& $0.020$ & $0.294$ 	& $(-0.662)_{\rm exp}$ \\  
EOGM ($g_A/g_V=1.25$), Engel--Vogel~\cite{Engel:1989ix} 
	&  $0.047$ & $0.2646$ 	& $(-0.662)_{\rm exp}$ \\ 
\hline
\hline 
$^{23}$Na~($L_J=P_{3/2}$) & 
~~~~~~~~$\langle {\bf S}_p \rangle$ & 
~~~~~~~~$\langle {\bf S}_n \rangle$ & ~~~~~~~~$\mu$ (in $\mu_N$) \\ \hline
ISPSM 
	& $1/2$ 	& 0		& $3.793$  \\ 
SM, Ressell-Dean~\cite{Ressell:1997kx}
	& 0.2477 	& 0.0198 	& 2.2196 \\ 
OGM, Ressell-Dean~\cite{Ressell:1997kx}      
        & 0.1566 	& 0.0    & (2.218)$_{\rm exp}$ \\ 
SM, Divari ar al.~\cite{Divari:2000dc} 
	& 0.2477 & 0.0199 & 2.22   \\ 
\hline
\hline 
$^{25}$Mg~($L_J=D_{5/2}$) & 
~~~~~~~~$\langle {\bf S}_p \rangle$ & 
~~~~~~~~$\langle {\bf S}_n \rangle$ & ~~~~~~~~$\mu$ (in $\mu_N$) \\ \hline
ISPSM 
	& 0 	& $1/2$ 	& $-1.913$  \\ 
OGM 
	& 0 	& $0.223$ 	& $(-0.855)_{\rm exp}$ \\ 
EOGM ($g_A/g_V=1$), Engel--Vogel~\cite{Engel:1989ix} 
	& $0.040$ & $0.376$ 	& $(-0.855)_{\rm exp}$ \\  
EOGM ($g_A/g_V=1.25$), Engel--Vogel~\cite{Engel:1989ix} 
	&  $0.073$ & $0.343$ 	& $(-0.855)_{\rm exp}$ \\ 
\hline
\hline 
$^{27}$Al~($L_J=D_{5/2}$) & 
~~~~~~~~$\langle {\bf S}_p \rangle$ & 
~~~~~~~~$\langle {\bf S}_n \rangle$ & 
~~~~~~~~$\mu$ (in $\mu_N$) \\ \hline
ISPSM, Ellis--Flores~\cite{Ellis:1988sh,Ellis:1991ef}
	& $1/2$ 	& 0 	  & $4.793$  \\ 
OGM, Engel--Vogel~\cite{Engel:1989ix} 
	& $0.25$ 	& 0 	  & $(3.642)_{\rm exp}$ \\ 
EOGM ($g_A/g_V=1$), Engel--Vogel~\cite{Engel:1989ix} 
	& $0.333$ 	& $0.043$ & $(3.642)_{\rm exp}$ \\ 
EOGM ($g_A/g_V=1.25$), Engel--Vogel~\cite{Engel:1989ix} 
	&  $0.304$ 	& $0.072$ & $(3.642)_{\rm exp}$ \\ 
SM, Engel et al.~\cite{Engel:1995gw}
	& $0.3430$ & $0.0296$ 
	& $3.584$  \\
\hline
\hline
\end{tabular} \end{center}
\end{table} 

\begin{table}[!ht] 
\caption{Zero momentum spin structure of light nuclei 
	($29<A<41$) in different models. 
\label{Nuclear.spin.main.table.29-41}}
\begin{center}
\begin{tabular}{lrrr}
\hline
\hline
$^{29}$Si~($L_J=S_{1/2}$) & 
~~~~~~~~$\langle {\bf S}_p \rangle$ & 
~~~~~~~~$\langle {\bf S}_n \rangle$ & 
~~~~~~~~$\mu$ (in $\mu_N$) \\ \hline
ISPSM, Ellis--Flores~\cite{Ellis:1988sh,Ellis:1991ef}
	& 0 & $1/2$ & $-1.913$  \\ 
OGM, Engel--Vogel~\cite{Engel:1989ix} 
	& 0 & $0.15$ & $(-0.555)_{\rm exp}$ \\ 
EOGM ($g_A/g_V=1$), Engel--Vogel~\cite{Engel:1989ix} 
	& $0.054$ & $0.204$ & $(-0.555)_{\rm exp}$ \\ 
EOGM ($g_A/g_V=1.25$), Engel--Vogel~\cite{Engel:1989ix} 
	&  $0.069$ & $0.189$ & $(-0.555)_{\rm exp}$ \\ 
SM, Ressell et al.~\cite{Ressell:1993qm}	
	&$-0.002$ & $0.13$ & $-0.50$ \\
SM, Divari et al.~\cite{Divari:2000dc} 
	&$-0.0019$ &0.1334 
	&$-0.50$ \\ 
\hline
\hline
$^{31}$P~($L_J=S_{1/2}$) & 
~~~~~~~~$\langle {\bf S}_p \rangle$ & 
~~~~~~~~$\langle {\bf S}_n \rangle$ & 
~~~~~~~~$\mu$ (in $\mu_N$) \\ \hline
ISPSM 
	& $0.5$ & 0 & $2.793 $  \\ 
OGM 
	& $0.138$ & 0 & $(1.132)_{\rm exp}$ \\ 
EOGM ($g_A/g_V=1$), Engel--Vogel~\cite{Engel:1989ix} 
	& $0.181$ & $0.032$ & $(1.132)_{\rm exp}$ \\  
EOGM ($g_A/g_V=1.25$), Engel--Vogel~\cite{Engel:1989ix} 
	&  $0.166$ & $0.047$ & $(1.132)_{\rm exp}$ \\ 
\hline 
\hline 
$^{35}$Cl~($L_J=D_{3/2}$) & 
~~~~~~~~$\langle {\bf S}_p \rangle$ & 
~~~~~~~~$\langle {\bf S}_n \rangle$ & 
~~~~~~~~$\mu$ (in $\mu_N$) \\
\hline
ISPSM, Ellis--Flores~\cite{Ellis:1988sh,Ellis:1991ef}
	&$-0.3$   &    0 &~~0.13	 \\
OGM, Engel--Vogel~\cite{Engel:1989ix} 
	&$-0.15$  &    0   & (0.822)$_{\rm exp}$	 \\
EOGM, Engel--Vogel~\cite{Engel:1989ix} 
	&$-0.094$  & $0.014$ & (0.822)$_{\rm exp}$	 \\	
SM, Pacheco-Strottman~\cite{Pacheco:1989jz}
	&$-0.059$  &$-0.011$ \\	
SM, Ressell et al.~\protect\cite{Ressell:1993qm} 	
	&~~$-0.051$&~~$-0.0088$ \\ 
\hline
\hline 
$^{39}$K~($L_J=D_{3/2}$) & 
~~~~~~~~$\langle {\bf S}_p \rangle$ & 
~~~~~~~~$\langle {\bf S}_n \rangle$ & 
~~~~~~~~$\mu$ (in $\mu_N$) \\ 
\hline
 ISPSM 
	& $-0.3$ & 0 &  $0.324$ \\ 
 OGM 
	& $-0.242$ &   0  & $(0.391)_{\rm exp}$	 \\
 EOGM ($g_A/g_V = 1.0$) Engel--Vogel~\cite{Engel:1989ix}
	  & $-0.196$ & $0.055$ &  $(0.391)_{\rm exp}$	 \\
 EOGM ($g_A/g_V = 1.25$) Engel--Vogel~\cite{Engel:1989ix}
	  & $-0.171$ & $0.030$ &  $(0.391)_{\rm exp}$	 \\
PT with Force I,  Engel et al.~\cite{Engel:1995gw} 
	& $-0.197$ & $0.051$ &    $0.420$ \\ 
PT with Force II, Engel et al.~\cite{Engel:1995gw} 
	& $-0.184$ & $0.054$ & $0.181$  \\ 
\hline
\hline
\end{tabular} \end{center}
\end{table} 

\begin{table}[!ht] 
\caption{Zero momentum spin structure of some nuclei 
	($45<A<73$) in different models. 
\label{Nuclear.spin.main.table.45-73}}
\begin{center}
\begin{tabular}{lrrr}
\hline
\hline
$^{47}$Ti~($L_J=F_{5/2}$)& ~~~~~~~~$\langle {\bf S}_p \rangle$ & 
~~~~~~~~~~~~~~~~$\langle {\bf S}_n \rangle$ & 
~~~~~~~~$\mu$ (in $\mu_N$) \\ \hline
ISPSM, Ellis--Flores~\cite{Ellis:1988sh,Ellis:1991ef}
	& 0 & $-0.357$ & $1.367$  \\ 
OGM, Engel--Vogel~\cite{Engel:1989ix} 
	& 0 & $0.21$ & $(-0.788)_{\rm exp}$ \\ 
\hline
\hline
$^{49}$Ti~($L_J=F_{7/2}$) & ~~~~~~~~$\langle {\bf S}_p \rangle$ & 
~~~~~~~~$\langle {\bf S}_n \rangle$ & ~~~~~~~~$\mu$ (in $\mu_N$) \\ \hline
ISPSM, Ellis--Flores~\cite{Ellis:1988sh,Ellis:1991ef}
	& 0 & $1/2$ & $-1.913$  \\ 
OGM, Engel--Vogel~\cite{Engel:1989ix} 
	& 0 & $0.29$ & $(-1.104)_{\rm exp}$ \\ 
\hline
\hline
$^{51}$V~($L_J=F_{7/2}$) & 
~~~~~~~~$\langle {\bf S}_p \rangle$ & 
~~~~~~~~$\langle {\bf S}_n \rangle$ & 
~~~~~~~~$\mu$ (in $\mu_N$) \\ \hline
ISPSM, Ellis--Flores~\cite{Ellis:1988sh,Ellis:1991ef}
	&$1/2$& 0 & $5.79$  \\ 
OGM, Engel--Vogel~\cite{Engel:1989ix} 
	&$0.36$ & 0 & $(5.149)_{\rm exp}$ \\ 
\hline 
\hline 
$^{55}$Mn~($L_J=F_{5/2}$) & 
~~~~~~~~$\langle {\bf S}_p \rangle$ & 
~~~~~~~~$\langle {\bf S}_n \rangle$ & 
~~~~~~~~$\mu$ (in $\mu_N$) \\
\hline
ISPSM, Ellis--Flores~\cite{Ellis:1991ef,Ellis:1993vh}
	&$-0.357$   &    0 & $5.79$	 \\
OGM, Ellis--Flores~\cite{Ellis:1991ef,Ellis:1993vh}
	&$0.264$  &    0   & $(3.453)_{\rm exp}$	 \\
\hline
\hline 
$^{59}$Co~($L_J=F_{7/2}$) & 
~~~~~~~~$\langle {\bf S}_p \rangle$ & 
~~~~~~~~$\langle {\bf S}_n \rangle$ & 
~~~~~~~~$\mu$ (in $\mu_N$) \\ 
\hline
ISPSM, Ellis--Flores~\cite{Ellis:1991ef,Ellis:1993vh}
	&$1/2$   &    0 & $5.79$	 \\
OGM, Ellis--Flores~\cite{Ellis:1991ef,Ellis:1993vh}
	&$0.25$  &    0   & $(4.627)_{\rm exp}$	 \\
\hline
\hline 
$^{67}$Zn~($L_J=F_{5/2}$) & 
~~~~~~~~$\langle {\bf S}_p \rangle$ & 
~~~~~~~~$\langle {\bf S}_n \rangle$ & 
~~~~~~~~$\mu$ (in $\mu_N$) \\ \hline
ISPSM, Ellis--Flores~\cite{Ellis:1988sh,Ellis:1991ef}
	& 0 & $-0.357$ & $1.367$  \\ 
OGM, Engel--Vogel~\cite{Engel:1989ix} 
	& 0 & $-0.23$ & $(0.875)_{\rm exp}$ \\ 
\hline
\hline 
$^{69}$Ga~($L_J=P_{3/2}$) & 
~~~~~~~~$\langle {\bf S}_p \rangle$ & 
~~~~~~~~$\langle {\bf S}_n \rangle$ & 
~~~~~~~~$\mu$ (in $\mu_N$) \\ 
\hline
ISPSM, Ellis--Flores~\cite{Ellis:1988sh,Ellis:1991ef}
	& $0.5$ & 0 &  $3.793$ \\ 
OGM, Engel--Vogel~\cite{Engel:1989ix} 
	&$0.11$&   0  & $(2.017)_{\rm exp}$	 \\
\hline
\hline
$^{71}$Ga~($L_J=P_{3/2}$) & 
~~~~~~~~$\langle {\bf S}_p \rangle$ & 
~~~~~~~~$\langle {\bf S}_n \rangle$ & 
~~~~~~~~$\mu$ (in $\mu_N$) \\ 
\hline
ISPSM, Ellis--Flores~\cite{Ellis:1988sh,Ellis:1991ef}
	& $0.5$ & 0 &  $3.793$ \\ 
OGM, Engel--Vogel~\cite{Engel:1989ix} 
	& $0.23$&   0  & $(2.562)_{\rm exp}$	 \\
\hline
\hline
\end{tabular} \end{center}
\end{table} 

\begin{table}[!ht] 
\caption{Zero momentum spin structure of some nuclei 
	($71<A<95$) in different models. 
\label{Dimitrov:1995gc:t2}
\label{Nuclear.spin.main.table.71-95}}
\begin{center}
\begin{tabular}{lrrr}
\hline\hline
$^{73}$Ge~($L_J=G_{9/2}$) & ~~~~~~~~$\langle {\bf S}_p \rangle$ & 
~~~~~~~~$\langle {\bf S}_n \rangle$ & ~~~~~~~~$\mu$ (in $\mu_N$) \\ \hline
ISPSM, Ellis--Flores~\cite{Ellis:1988sh,Ellis:1991ef}
	&    0	  & $0.5$		& $-1.913$ \\ 
OGM, Engel--Vogel~\cite{Engel:1989ix} 	
	&    0	  & $0.23$ 	&$(-0.879)_{\rm exp}$ \\ 
IBFM, Iachello et al.~\cite{Iachello:1991ut} and \cite{Ressell:1993qm}
	&$-0.009$ & $0.469$ &$-1.785$\\ 
IBFM (quenched), 
	Iachello et al.~\cite{Iachello:1991ut} and \cite{Ressell:1993qm}
	&$-0.005$  & $0.245$ &$(-0.879)_{\rm exp}$ \\
TFFS, Nikolaev--Klapdor-Kleingrothaus, \cite{Nikolaev:1993dd} 
	&$0$   & $0.34$ & --- \\ 
SM (small), Ressell et al.~\cite{Ressell:1993qm} 
	&$0.005$   & $0.496$ &$-1.468$ \\ 
SM (large), Ressell et al.~\cite{Ressell:1993qm} 
	&$0.011$   & $0.468$ &$-1.239$ \\ 
SM (large, quenched), Ressell et al.~\cite{Ressell:1993qm} 
	&$0.009$   & $0.372$ &$(-0.879)_{\rm exp}$ \\ 
``Hybrid'' SM, Dimitrov et al.~\cite{Dimitrov:1995gc}           
	& $0.030$ & $0.378$ & $-0.920$ \\ 
\hline
\hline
$^{75}$As~($L_J=P_{3/2}$) & 
~~~~~~~~$\langle {\bf S}_p \rangle$ & 
~~~~~~~~$\langle {\bf S}_n \rangle$ & 
~~~~~~~~$\mu$ (in $\mu_N$) \\ \hline
ISPSM, Ellis--Flores~\cite{Ellis:1988sh,Ellis:1991ef}
	& $0.5$ & 0 &  $3.793$ \\ 
OGM, Engel--Vogel~\cite{Engel:1989ix} 
	& $-0.01$&   0  & $(1.439)_{\rm exp}$	 \\
\hline
\hline
$^{79}$Br~($L_J=P_{3/2}$) & 
~~~~~~~~$\langle {\bf S}_p \rangle$ & 
~~~~~~~~$\langle {\bf S}_n \rangle$ & 
~~~~~~~~$\mu$ (in $\mu_N$) \\ \hline
ISPSM, Ellis--Flores~\cite{Ellis:1988sh,Ellis:1991ef}
	& $0.5$ & 0 &  $3.793$ \\ 
OGM, Engel--Vogel~\cite{Engel:1989ix} 
	& $0.13$&   0  & $(2.106)_{\rm exp}$	 \\
\hline
\hline
$^{81}$Br~($L_J=P_{3/2}$) & 
~~~~~~~~$\langle {\bf S}_p \rangle$ & 
~~~~~~~~$\langle {\bf S}_n \rangle$ & 
~~~~~~~~$\mu$ (in $\mu_N$) \\ \hline
ISPSM, Ellis--Flores~\cite{Ellis:1988sh,Ellis:1991ef}
	& $0.5$ & 0 &  $3.793$ \\ 
OGM, Engel--Vogel~\cite{Engel:1989ix} 
	& $0.17$&   0  & $(2.271)_{\rm exp}$	 \\
\hline
\hline
$^{91}$Zr~($L_J=D_{5/2}$) & 
~~~~~~~~$\langle {\bf S}_p \rangle$ & 
~~~~~~~~$\langle {\bf S}_n \rangle$ & 
~~~~~~~~$\mu$ (in $\mu_N$) \\ \hline
ISPSM, Ellis--Flores~\cite{Ellis:1988sh,Ellis:1991ef}
 	& $0$ & $0.5$ &  $-1.913$ \\ 
OGM, Engel--Vogel~\cite{Engel:1989ix} 
	& $0$&   $0.34$  & $(-1.304)_{\rm exp}$	 \\
\hline
\hline 
$^{93}$Nb~($L_J=G_{9/2}$) &
~~~~~~~~$\langle {\bf S}_p \rangle$ & 
~~~~~~~~$\langle {\bf S}_n \rangle$ & 
~~~~~~~~$\mu$ (in $\mu_N$) \\ \hline
ISPSM, Ellis--Flores~\cite{Ellis:1988sh,Ellis:1991ef}
	& $0.5$ & 0 &  $6.793$ \\ 
OGM, Engel--Vogel~\cite{Engel:1989ix} 
	& $0.36$&   0  & $(6.171)_{\rm exp}$	 \\
SM (large), Engel et al.~\cite{Engel:1992qb}
	   &0.48     & 0.04 & $6.36$ \\ 
SM (small), Engel et al.~\cite{Engel:1992qb}
	   &0.46     & 0.08 & $5.88$ \\ 
\hline\hline
\end{tabular} \end{center}
\end{table} 

\begin{table}[!ht] 
\caption{Zero momentum spin structure of some nuclei 
	($95<A<115$) in different models. 
\label{Nuclear.spin.main.table.95-115}}
\begin{center}
\begin{tabular}{lrrr}
\hline
\hline
$^{99}$Ru~($L_J=D_{5/2}$) & 
~~~~~~~~$\langle {\bf S}_p \rangle$ & 
~~~~~~~~$\langle {\bf S}_n \rangle$ & 
~~~~~~~~$\mu$ (in $\mu_N$) \\ \hline
ISPSM, Ellis--Flores~\cite{Ellis:1991ef,Ellis:1993vh}
	& 0 & $1/2$ & $-1.913$  \\ 
OGM, Ellis--Flores~\cite{Ellis:1991ef,Ellis:1993vh}
	& 0 & $0.17$ & $(-0.6381)_{\rm exp}$ \\ 
\hline
\hline
$^{101}$Ru~($L_J=D_{5/2}$) & 
~~~~~~~~$\langle {\bf S}_p \rangle$ & 
~~~~~~~~$\langle {\bf S}_n \rangle$ & 
~~~~~~~~$\mu$ (in $\mu_N$) \\ \hline
ISPSM, Ellis--Flores~\cite{Ellis:1991ef,Ellis:1993vh}
	& 0 & $1/2$ & $-1.913$  \\ 
OGM, Ellis--Flores~\cite{Ellis:1991ef,Ellis:1993vh}
	& 0 & $0.19$ & $(-0.719)_{\rm exp}$ \\ 
\hline
\hline
$^{107}$Ag~($L_J=P_{1/2}$) & 
~~~~~~~~$\langle {\bf S}_p \rangle$ & 
~~~~~~~~$\langle {\bf S}_n \rangle$ & 
~~~~~~~~$\mu$ (in $\mu_N$) \\ \hline
ISPSM, Ellis--Flores~\cite{Ellis:1988sh,Ellis:1991ef}
	&$-0.167$& 0 & $-0.264[-0.07]$  \\ 
OGM, Engel--Vogel~\cite{Engel:1989ix} 
	&$-0.13$ & 0 & $(-0.114)_{\rm exp}$ \\ 
\hline 
\hline 
$^{109}$Ag~($L_J=P_{1/2}$) & 
~~~~~~~~$\langle {\bf S}_p \rangle$ & 
~~~~~~~~$\langle {\bf S}_n \rangle$ & 
~~~~~~~~$\mu$ (in $\mu_N$) \\ \hline
ISPSM, Ellis--Flores~\cite{Ellis:1988sh,Ellis:1991ef}
	&$-0.167$   &    0 & $-0.264[-0.07]$	 \\
OGM, Ellis--Flores~\cite{Ellis:1991ef,Ellis:1993vh}
	&$-0.14$  &    0   & $(-0.131)_{\rm exp}$	 \\
\hline
\hline 
$^{111}$Cd~($L_J=S_{1/2}$) & 
~~~~~~~~$\langle {\bf S}_p \rangle$ & 
~~~~~~~~$\langle {\bf S}_n \rangle$ & 
~~~~~~~~$\mu$ (in $\mu_N$) \\ \hline
ISPSM, Ellis--Flores~\cite{Ellis:1988sh,Ellis:1991ef}
	& 0 & $1/2$ & $-1.913$  \\ 
OGM, Engel--Vogel~\cite{Engel:1989ix} 
	& 0 & $0.16$ & $(-0.595)_{\rm exp}$ \\ 
\hline
\hline 
$^{113}$Cd~($L_J=S_{1/2}$) & 
~~~~~~~~$\langle {\bf S}_p \rangle$ & 
~~~~~~~~$\langle {\bf S}_n \rangle$ & 
~~~~~~~~$\mu$ (in $\mu_N$) \\ \hline
ISPSM, Ellis--Flores~\cite{Ellis:1988sh,Ellis:1991ef}
	& 0 & $1/2$ & $-1.913$  \\ 
OGM, Engel--Vogel~\cite{Engel:1989ix} 
	& 0 & $0.16$ & $(-0.622)_{\rm exp}$ \\ 
IBFM, Iachello et al.~\cite{Iachello:1991ut}
	&$-0.001$ & $0.488$ & --- \\
IBFM (quenched), 
	Iachello et al.~\cite{Iachello:1991ut}
	&$-0.0$  & $0.162$ & --- \\
TFFS, Nikolaev--Klapdor-Kleingrothaus, \cite{Nikolaev:1993dd} 
	&$0$   & $0.175$ & --- \\ 
\hline
\hline 
$^{115}$Cd~($L_J=S_{1/2}$) & 
~~~~~~~~$\langle {\bf S}_p \rangle$ & 
~~~~~~~~$\langle {\bf S}_n \rangle$ & 
~~~~~~~~$\mu$ (in $\mu_N$) \\ \hline
ISPSM, Nikolaev--Klapdor-Kleingrothaus, \cite{Nikolaev:1993dd} 
	& 0 & $1/2$ &  --- \\ 
IBFM, Iachello et al.~\cite{Iachello:1991ut}
	&$-0.001$ & $0.488$ & --- \\
IBFM (quenched), 
	Iachello et al.~\cite{Iachello:1991ut}
	&$-0.0$  & $0.168$ &  --- \\
TFFS, Nikolaev--Klapdor-Kleingrothaus, \cite{Nikolaev:1993dd} 
	&$0$   & $0.195$ & --- \\ 
OGM 
	& 0 & $0.169$ & $(-0.6388)_{\rm exp}$ \\ 
\hline
\hline
\end{tabular} \end{center}
\end{table} 

\begin{table}[!ht] 
\caption{Zero momentum spin structure of heavy nuclei 
	($114<A<125$) in different models. 
	Calculations of the magnetic moment using 
	effective $g$-factors are given in curly brackets.
\label{Nuclear.spin.main.table.114-125}}
\begin{center}
\begin{tabular}{lrrr}
\hline
\hline
$^{115}$Sn~($L_J=S_{1/2}$) & 
~~~~~~~~$\langle {\bf S}_p \rangle$ & 
~~~~~~~~$\langle {\bf S}_n \rangle$ & 
~~~~~~~~$\mu$ (in $\mu_N$) \\ \hline
ISPSM, Ellis--Flores~\cite{Ellis:1988sh,Ellis:1991ef}
	& 0 & $1/2$ &  $-1.913$ \\ 
OGM, Engel--Vogel~\cite{Engel:1989ix} 
	& 0 & $0.24$ & $(-0.919)_{\rm exp}$ \\ 
\hline
\hline
$^{117}$Sn~($L_J=S_{1/2}$) & 
~~~~~~~~$\langle {\bf S}_p \rangle$ & 
~~~~~~~~$\langle {\bf S}_n \rangle$ & 
~~~~~~~~$\mu$ (in $\mu_N$) \\ \hline
ISPSM, Ellis--Flores~\cite{Ellis:1988sh,Ellis:1991ef}
	& 0 & $1/2$ &  $-1.913$ \\ 
OGM, Engel--Vogel~\cite{Engel:1989ix} 
	& 0 & $0.126$ & $(-1.001)_{\rm exp}$ \\ 
\hline
\hline
$^{121}$Sb~($L_J=D_{5/2}$) & 
~~~~~~~~$\langle {\bf S}_p \rangle$ & 
~~~~~~~~$\langle {\bf S}_n \rangle$ & 
~~~~~~~~$\mu$ (in $\mu_N$) \\ \hline
ISPSM, Ellis--Flores~\cite{Ellis:1988sh,Ellis:1991ef}
	&$0.5$& 0 & $4.793$  \\ 
OGM, Ellis--Flores~\cite{Ellis:1991ef,Ellis:1993vh}
	&$0.188$ & 0 & $(3.363)_{\rm exp}$ \\ 
\hline
\hline
$^{123}$Sb~($L_J=G_{7/2}$) & 
~~~~~~~~$\langle {\bf S}_p \rangle$ & 
~~~~~~~~$\langle {\bf S}_n \rangle$ & 
~~~~~~~~$\mu$ (in $\mu_N$) \\ \hline
ISPSM, Ellis--Flores~\cite{Ellis:1988sh,Ellis:1991ef}
	&$-0.389$& 0 & $1.717$  \\ 
OGM, Ellis--Flores~\cite{Ellis:1991ef,Ellis:1993vh}
	&$-0.207$ & 0 & $(2.550)_{\rm exp}$ \\ 
\hline
\hline
$^{123}$Te~($L_J=S_{1/2}$) & 
~~~~~~~~$\langle {\bf S}_p \rangle$ & 
~~~~~~~~$\langle {\bf S}_n \rangle$ & 
~~~~~~~~$\mu$ (in $\mu_N$) \\ \hline
ISPSM, Ellis--Flores~\cite{Ellis:1988sh,Ellis:1991ef}
	& 0 & $1/2$ & $-1.913$ \\
IBFM, Iachello et al.~\cite{Iachello:1991ut}
	&$-0.000$ & $0.491$ & --- \\
IBFM (quenched), 
	Iachello et al.~\cite{Iachello:1991ut}
	&$-0.000$  & $0.192$ &  --- \\
TFFS, Nikolaev--Klapdor-Kleingrothaus, \cite{Nikolaev:1993dd} 
	&    & $0.21$ & --- \\ 
OGM 
	& 0 & $0.192$ & $(-0.737)_{\rm exp}$ \\ 
\hline
\hline
$^{125}$Te~($L_J=S_{1/2}$) & 
~~~~~~~~$\langle {\bf S}_p \rangle$ & 
~~~~~~~~$\langle {\bf S}_n \rangle$ & 
~~~~~~~~~~~~~~~~$\mu$ (in $\mu_N$) \\ \hline
ISPSM, Ellis--Flores~\cite{Ellis:1988sh,Ellis:1991ef}
	& 0 & $1/2$ & $-1.913$ \\   
OGM, Engel--Vogel~\cite{Engel:1989ix} 
	& 0.0  & $0.23$ & $(-0.889)_{\rm exp}$ \\ 
IBFM, Iachello et al.~\cite{Iachello:1991ut}
	&$-0.0008$ & $0.499$ & $(-0.889)_{\rm exp}$ \\
IBFM (quenched), 
	Iachello et al.~\cite{Iachello:1991ut}
	&$-0.0004$  & $0.231$ & $(-0.889)_{\rm exp}$ \\
TFFS, Nikolaev--Klapdor-Kleingrothaus, \cite{Nikolaev:1993dd} 
	&   & $0.22$ & --- \\ 
SM (Bonn A), Ressell--Dean~\cite{Ressell:1997kx} 
        & $0.001$ &$0.287$& $-1.015~\{-0.749\}_{\rm eff}$ \\ 
SM (Nijmegen II), Ressell--Dean~\cite{Ressell:1997kx} 
	&$-0.0003$&$0.323$ &$-1.134~\{-0.824\}_{\rm eff}$ \\ 
\hline
\hline 
\end{tabular} \end{center}
\end{table} 

\begin{table}[!ht] 
\caption{Zero momentum spin structure of heavy nuclei 
	($125<A<127$) in different models. 
	Calculations of the magnetic moment using 
	effective $g$-factors are given in curly brackets.
\label{Nuclear.spin.main.table.125-127}}
\begin{center}
\begin{tabular}{lrrr}
\hline
\hline
$^{125}$I~($L_J=D_{5/2}$) & 
~~~~~~~~$\langle {\bf S}_p \rangle$ & 
~~~~~~~~$\langle {\bf S}_n \rangle$ & 
~~~~~~~~$\mu$ (in $\mu_N$) \\ \hline
ISPSM, Engel--Vogel~\cite{Engel:1989ix} 
	&$1/2$ & 0 & $4.793$ \\
IBFM, Iachello et al.~\cite{Iachello:1991ut}
	&$0.460$ & $0.005$ &  \\
IBFM (quenched), 
	Iachello et al.~\cite{Iachello:1991ut}
	&$0.159$  & $0.002$ &  \\
TFFS, Nikolaev--Klapdor-Kleingrothaus, \cite{Nikolaev:1993dd} 
	& $0.18$ &  &  \\ 
OGM 
	& $0.07$ & 0 & $(2.821)_{\rm exp}$ \\ 
\hline
\hline
$^{127}$I~($L_J=D_{5/2}$) & 
~~~~~~~~$\langle {\bf S}_p \rangle$ & ~~~~~~~~$\langle {\bf S}_n \rangle$ & 
~~~~~~~~~~~~~~~~$\mu$ (in $\mu_N$) \\ \hline
ISPSM, Ellis--Flores~\cite{Ellis:1991ef,Ellis:1993vh}
	&$1/2$ & 0 &  $4.793$ \\ 
OGM, Engel--Vogel~\cite{Engel:1989ix} 
	&$0.07$ & 0 & $(2.813)_{\rm exp}$ \\ 
IBFM, Iachello et al.~\cite{Iachello:1991ut}
	&$0.464$ & $0.010$ & $(2.813)_{\rm exp}$ \\ 
IBFM (quenched), 
	Iachello et al.~\cite{Iachello:1991ut}
	&$0.154$  & $0.003$ & $(2.813)_{\rm exp}$ \\ 
TFFS, Nikolaev--Klapdor-Kleingrothaus, \cite{Nikolaev:1993dd} 
	& $0.15$ & 0 & --- \\ 
SM (Bonn A), Ressell--Dean~\cite{Ressell:1997kx} 
	& $0.309$ & $0.075$ & $2.775~\{2.470\}_{\rm eff}$  \\ 
SM (Nijmegen II), Ressell--Dean~\cite{Ressell:1997kx} 
	& $0.354$ & $0.064$ & $3.150~\{2.7930\}_{\rm eff}$ \\ 
\hline
\hline 
\end{tabular} \end{center}
\end{table} 

\begin{table}[!ht] 
\caption{Zero momentum spin structure of heavy nuclei 
	($128<A<133$) in different models. 
	Calculations of the magnetic moment using 
	effective $g$-factors are given in curly brackets.
\label{Nuclear.spin.main.table.128-133}}
\begin{center}
\begin{tabular}{lrrr}
\hline
\hline
$^{129}$Xe~($L_J=S_{1/2}$) & 
~~~~~~~~$\langle {\bf S}_p \rangle$ & 
~~~~~~~~$\langle {\bf S}_n \rangle$ & 
~~~~~~~~~~~$\mu$ (in $\mu_N$) \\ \hline
ISPSM, Ellis--Flores~\cite{Ellis:1988sh,Ellis:1991ef}
	& 0 & $1/2$ & $-1.913$ \\   
OGM, Engel--Vogel~\cite{Engel:1989ix} 
	& 0.0  & $0.2$ & $(-0.778)_{\rm exp}$ \\ 
IBFM, Iachello et al.~\cite{Iachello:1991ut}
	&$-0.000$ & $0.430$ & $(-0.778)_{\rm exp}$ \\
IBFM (quenched), Iachello et al.~\cite{Iachello:1991ut}
	&$-0.000$  & $0.200$ & $(-0.788)_{\rm exp}$ \\
TFFS, Nikolaev--Klapdor-Kleingrothaus, \cite{Nikolaev:1993dd} 
	&   & $0.25$ & --- \\ 
SM (Bonn A), Ressell--Dean~\cite{Ressell:1997kx} 
	& $0.028$& $0.359$& $-0.983~\{-0.634\}_{\rm eff}$ \\ 
SM (Nijmegen II), Ressell--Dean~\cite{Ressell:1997kx} 
	& $0.0128$& $0.300$ & $-0.701\{-0.379\}_{\rm eff}$ \\ 
\hline
\hline
$^{131}$Xe~($L_J=D_{3/2}$) & 
~~~~~~~~$\langle {\bf S}_p \rangle$ & 
~~~~~~~~$\langle {\bf S}_n \rangle$ & 
~~~~~~~~$\mu$ (in $\mu_N$) \\ \hline 
ISPSM, Ellis--Flores~\cite{Ellis:1988sh,Ellis:1991ef}
	& 0 & $-0.3$ & $1.148$ \\
OGM, Engel--Vogel~\cite{Engel:1989ix} 
	& 0.0  & $-0.18$ 	& $(0.692)_{\rm exp}$ \\ 
IBFM, Iachello et al.~\cite{Iachello:1991ut}
	&$0.000$ & $-0.280$ 	& $(0.692)_{\rm exp}$ \\
IBFM (quenched), Iachello et al.~\cite{Iachello:1991ut}
	&$0.000$  & $-0.168$ 	& $(0.692)_{\rm exp}$ \\
TFFS, Nikolaev--Klapdor-Kleingrothaus, \cite{Nikolaev:1993dd} 
	&   & $-0.186$ & --- \\ 
SM (Bonn A), Ressell--Dean~\cite{Ressell:1997kx} 
	& $-0.009$ & $-0.227$ & $0.980~\{0.637\}_{\rm eff}$  \\ 
SM (Nijmegen II), Ressell--Dean~\cite{Ressell:1997kx} 
	& $-0.012$ & $-0.217$ & $0.979~\{0.347\}_{\rm eff}$ \\  
QTDA, Engel~\cite{Engel:1991wq} 
	& $-0.041$ & $-0.236$ & 0.70 \\  
\hline
\hline 
$^{133}$Xe~($L_J=D_{3/2}$) & 
~~~~~~~~$\langle {\bf S}_p \rangle$ & 
~~~~~~~~$\langle {\bf S}_n \rangle$ & 
~~~~~~~~$\mu$ (in $\mu_N$) \\ \hline 
ISPSM, Ellis--Flores~\cite{Ellis:1988sh,Ellis:1991ef}
	& 0 & $-0.3$ & $1.148$ \\ 
IBFM, Iachello et al.~\cite{Iachello:1991ut}
	&$0.000$ & $-0.257$ &  \\
IBFM (quenched), Iachello et al.~\cite{Iachello:1991ut}
	&$0.000$  & $-0.176$ &  \\
TFFS, Nikolaev--Klapdor-Kleingrothaus, \cite{Nikolaev:1993dd} 
	&   & $-0.201$ &  \\ 
OGM 
	& 0.0  & $-0.213$ & $(0.813)_{\rm exp}$ \\ 
\hline
\hline 
\end{tabular} \end{center}
\end{table} 

\begin{table}[!ht] 
\caption{Zero momentum spin structure of heavy nuclei 
	($133<A<141$) in different models. 
\label{Nuclear.spin.main.table.113-141}}
\begin{center}
\begin{tabular}{lrrr}
\hline
\hline
$^{133}$Cs~($L_J=G_{7/2}$) & 
~~~~~~~~$\langle {\bf S}_p \rangle$ & 
~~~~~~~~$\langle {\bf S}_n \rangle$ & 
~~~~~~~~$\mu$ (in $\mu_N$) \\ \hline 
ISPSM, Ellis--Flores~\cite{Ellis:1988sh,Ellis:1991ef}
	& $-0.389$ & 0 & $1.717$ \\ 
OGM, Engel--Vogel~\cite{Engel:1989ix} 
	& $-0.20$ & 0 & $(2.582)_{\rm exp}$ \\ 
IBFM, Iachello et al.~\cite{Iachello:1991ut}
	&$-0.370$  & $0.003$ &  \\
IBFM (quenched), Iachello et al.~\cite{Iachello:1991ut}
	&$-0.225$  & $0.002$ &  \\
TFFS, Nikolaev--Klapdor-Kleingrothaus, \cite{Nikolaev:1993dd} 
	& $-0.230$ &  &  \\ 
\hline
\hline
$^{135}$Cs~($L_J=G_{7/2}$) & 
~~~~~~~~$\langle {\bf S}_p \rangle$ & 
~~~~~~~~$\langle {\bf S}_n \rangle$ & 
~~~~~~~~$\mu$ (in $\mu_N$) \\ \hline 
ISPSM, Ellis--Flores~\cite{Ellis:1988sh,Ellis:1991ef}
	& $-0.389$ & 0 & $1.717$ \\ 
OGM 
	& $-0.167$ & 0 & $(2.734)_{\rm exp}$ \\ 
IBFM, Iachello et al.~\cite{Iachello:1991ut}
	&$-0.373$  & $0.002$ &  \\
IBFM (quenched), Iachello et al.~\cite{Iachello:1991ut}
	&$-0.201$  & $0.001$ &  \\
TFFS, Nikolaev--Klapdor-Kleingrothaus, \cite{Nikolaev:1993dd} 
	& $-0.199$ &  &  \\ 
\hline
\hline
$^{135}$Ba~($L_J=D_{3/2}$) & 
~~~~~~~~$\langle {\bf S}_p \rangle$ & 
~~~~~~~~$\langle {\bf S}_n \rangle$ & 
~~~~~~~~$\mu$ (in $\mu_N$) \\ \hline 
ISPSM, Ellis--Flores~\cite{Ellis:1988sh,Ellis:1991ef}
	& 0 & $-0.30 $ & $1.148 $  \\ 
OGM 
	& 0 & $-0.219$ & $(0.838)_{\rm exp}$ \\ 
IBFM, Iachello et al.~\cite{Iachello:1991ut}
	& $-0.007$& $-0.226$ &  \\
IBFM (quenched), Iachello et al.~\cite{Iachello:1991ut}
 	& $-0.004$  & $-0.145$ &  \\
TFFS, Nikolaev--Klapdor-Kleingrothaus, \cite{Nikolaev:1993dd} 
	&  & $-0.18$ &  \\ 
\hline
\hline
$^{137}$La~($L_J=G_{7/2}$) & 
~~~~~~~~$\langle {\bf S}_p \rangle$ & 
~~~~~~~~$\langle {\bf S}_n \rangle$ & 
~~~~~~~~$\mu$ (in $\mu_N$) \\ \hline 
ISPSM, Ellis--Flores~\cite{Ellis:1988sh,Ellis:1991ef}
	& $-0.389$ & 0 & $1.717$ \\ 
OGM 
	& $-0.176 $ & 0 & $(2.695)_{\rm exp}$ \\ 
IBFM, Iachello et al.~\cite{Iachello:1991ut}
	&$-0.386$  & $0.0006$ &  \\
IBFM (quenched), Iachello et al.~\cite{Iachello:1991ut}
	&$-0.212$  & $0.0003$ &  \\
\hline
\hline
$^{139}$La~($L_J=G_{7/2}$) & 
~~~~~~~~$\langle {\bf S}_p \rangle$ & 
~~~~~~~~$\langle {\bf S}_n \rangle$ & 
~~~~~~~~$\mu$ (in $\mu_N$) \\ \hline 
ISPSM, Ellis--Flores~\cite{Ellis:1988sh,Ellis:1991ef}
	& $-0.389$ & 0 & $1.717$ \\ 
OGM, Engel--Vogel~\cite{Engel:1989ix} 
	& $-0.16$ & 0 & $(2.783)_{\rm exp}$ \\ 
\hline
\hline
\end{tabular} \end{center}
\end{table} 

\begin{table}[!ht] 
\caption{Zero momentum spin structure of heavy nuclei 
	($143<A<205$) in different models. 
\label{Nuclear.spin.main.table.143-205}}
\begin{center}
\begin{tabular}{lrrr}
\hline
\hline
$^{155}$Gd~($L_J=P_{3/2}$) & 
~~~~~~~~$\langle {\bf S}_p \rangle$ & 
~~~~~~~~$\langle {\bf S}_n \rangle$ & 
~~~~~~~~$\mu$ (in $\mu_N$) \\ \hline 
ISPSM, 
	Ellis--Flores~\cite{Ellis:1991ef,Ellis:1993vh}
	& 0 & $0.5$ & $-1.913$  \\ 
OGM, Ellis--Flores~\cite{Ellis:1991ef,Ellis:1993vh}
	& 0 & $0.07$ & $(-0.259)_{\rm exp}$ \\ 
\hline
\hline
$^{157}$Gd~($L_J=P_{3/2}$) & 
~~~~~~~~$\langle {\bf S}_p \rangle$ & 
~~~~~~~~$\langle {\bf S}_n \rangle$ & 
~~~~~~~~$\mu$ (in $\mu_N$) \\ \hline 
ISPSM, 
	Ellis--Flores~\cite{Ellis:1991ef,Ellis:1993vh}
	& 0 & $0.5$ & $-1.913$  \\ 
OGM, Ellis--Flores~\cite{Ellis:1991ef,Ellis:1993vh}
	& 0 & $0.09$ & $(-0.340)_{\rm exp}$ \\ 
\hline
\hline
$^{183}$W~($L_J=P_{1/2}$) & 
~~~~~~~~$\langle {\bf S}_p \rangle$ & 
~~~~~~~~$\langle {\bf S}_n \rangle$ & 
~~~~~~~~$\mu$ (in $\mu_N$) \\ \hline 
ISPSM, 
	Ellis--Flores~\cite{Ellis:1991ef,Ellis:1993vh}
	& 0 & $-0.17$ & $0.638$  \\ 
OGM, Ellis--Flores~\cite{Ellis:1991ef,Ellis:1993vh}
	& 0 & $-0.03$ & $(0.118)_{\rm exp}$ \\ 
\hline
\hline
$^{191}$Ir~($L_J=D_{3/2}$) & 
~~~~~~~~$\langle {\bf S}_p \rangle$ & 
~~~~~~~~$\langle {\bf S}_n \rangle$ & 
~~~~~~~~$\mu$ (in $\mu_N$) \\ \hline 
ISPSM, Ellis--Flores~\cite{Ellis:1991ef,Ellis:1993vh}
	& $-0.30$ & 0 & $0.148$ \\ 
OGM, Ellis--Flores~\cite{Ellis:1991ef,Ellis:1993vh}
	& $-0.295$ & 0 & $(0.151)_{\rm exp}$ \\ 
\hline
\hline
$^{193}$Ir~($L_J=D_{3/2}$) & 
~~~~~~~~$\langle {\bf S}_p \rangle$ & 
~~~~~~~~$\langle {\bf S}_n \rangle$ & 
~~~~~~~~$\mu$ (in $\mu_N$) \\ \hline 
ISPSM, Ellis--Flores~\cite{Ellis:1991ef,Ellis:1993vh}
	& $-0.30$ & 0 & $0.148$ \\ 
OGM, Ellis--Flores~\cite{Ellis:1991ef,Ellis:1993vh}
	& $-0.292$ & 0 & $(0.164)_{\rm exp}$ \\ 
\hline
\hline
$^{199}$Hg~($L_J=P_{1/2}$) & 
~~~~~~~~$\langle {\bf S}_p \rangle$ & 
~~~~~~~~$\langle {\bf S}_n \rangle$ & 
~~~~~~~~$\mu$ (in $\mu_N$) \\ \hline 
ISPSM, Ellis--Flores~\cite{Ellis:1988sh,Ellis:1991ef}
	& 0 & $-0.17$ & $0.638$  \\ 
OGM, Engel--Vogel~\cite{Engel:1989ix} 
	& 0 & $-0.13$ & $(0.506)_{\rm exp}$ \\ 
\hline
\hline
$^{201}$Hg~($L_J=P_{3/2}$) & 
~~~~~~~~$\langle {\bf S}_p \rangle$ & 
~~~~~~~~$\langle {\bf S}_n \rangle$ & 
~~~~~~~~$\mu$ (in $\mu_N$) \\ \hline 
ISPSM, 
	Ellis--Flores~\cite{Ellis:1991ef,Ellis:1993vh}
	& 0 & $0.5$ & $-1.913$  \\ 
OGM, Ellis--Flores~\cite{Ellis:1991ef,Ellis:1993vh}
	& 0 & $0.146$ & $(-0.560)_{\rm exp}$ \\ 
\hline
\hline
$^{203}$Tl~($L_J=S_{1/2}$) & 
~~~~~~~~$\langle {\bf S}_p \rangle$ & 
~~~~~~~~$\langle {\bf S}_n \rangle$ & 
~~~~~~~~$\mu$ (in $\mu_N$) \\ \hline 
ISPSM, Ellis--Flores~\cite{Ellis:1988sh,Ellis:1991ef}
	& $0.50$ & 0 & $2.793$ \\ 
OGM, Engel--Vogel~\cite{Engel:1989ix} 
	& $0.24$ & 0 & $(1.662)_{\rm exp}$ \\ 
\hline
\hline
$^{205}$Tl~($L_J=S_{1/2}$) & 
~~~~~~~~$\langle {\bf S}_p \rangle$ & 
~~~~~~~~$\langle {\bf S}_n \rangle$ & 
~~~~~~~~$\mu$ (in $\mu_N$) \\ \hline 
ISPSM, Ellis--Flores~\cite{Ellis:1988sh,Ellis:1991ef}
	& $0.50$ & 0 & $2.793$ \\ 
OGM, Engel--Vogel~\cite{Engel:1989ix} 
	& $0.25$ & 0 & $(1.638)_{\rm exp}$ \\ 
\hline
\hline
\end{tabular} \end{center}
\end{table} 

\begin{table}[!ht] 
\caption{Zero momentum spin structure of heavy nuclei 
	($A<209$) in different models. 
\label{Nuclear.spin.main.table.209}}
\begin{center}
\begin{tabular}{lrrr}
\hline
\hline
$^{207}$Pb~($L_J=P_{1/2}$) & 
~~~~~~~~$\langle {\bf S}_p \rangle$ & 
~~~~~~~~$\langle {\bf S}_n \rangle$ & 
~~~~~~~~$\mu$ (in $\mu_N$) \\ \hline 
ISPSM, Ellis--Flores~\cite{Ellis:1988sh,Ellis:1991ef}
	& 0 & $-0.17$ & $0.638$  \\ 
OGM, Engel--Vogel~\cite{Engel:1989ix} 
	& 0 & $-0.15$ & $(0.593)_{\rm exp}$ \\ 
SM, Kosmas--Vergados~\cite{Kosmas:1997jm,Vergados:1996hs} 
	&$-0.010$ & $-0.149$ &  \\ 
\hline
\hline
$^{209}$Bi~($L_J=H_{9/2}$) & 
~~~~~~~~$\langle {\bf S}_p \rangle$ & 
~~~~~~~~$\langle {\bf S}_n \rangle$ & 
~~~~~~~~$\mu$ (in $\mu_N$) \\ \hline 
ISPSM, Ellis--Flores~\cite{Ellis:1991ef,Ellis:1993vh}
	& $-0.41$ & 0 & $2.63$ \\ 
OGM, Ellis--Flores~\cite{Ellis:1991ef,Ellis:1993vh}
	& $-0.085$ & 0 & $(4.111)_{\rm exp}$ \\ 
\hline
\hline
\end{tabular} \end{center}
\end{table} 

\bigskip

\clearpage

\section{Appendix A}

\def\WT{\widetilde}
\def\half{{1\over 2}}
\def\sq2{\sqrt{2}}
\def\CN{{\cal N}}
\def\sb{\sin\!\beta}
\def\cb{\cos\!\beta}
\def\tb{\tan\!\beta}
\def\sw{\sin\!\theta^{}_W}
\def\cw{\cos\!\theta^{}_W}
\def\tw{\tan\!\theta^{}_W}
\def\BM{\boldmath}
\newcommand{\mchi}{\mbox{$m_{\chi}$}}

\subsection{Elements of nuclear structure calculations}
	The transverse electric ${\cal T}^{el5}(q)$ 
	and longitudinal ${\cal L}^5(q)$ multipole projections of the
	axial vector current operator as well as the scalar 
	function ${\cal C}_L(q)$ are given by
\cite{Engel:1992bf,Engel:1991wq,Ressell:1993qm}:
\begin{eqnarray}
{\cal T}^{el5}_L(q) 
        &=& \frac{1}{\sqrt{2L+1}}\sum_i\frac{a_0 +a_1\tau^i_3}{2}
                 \Bigl[
		-\sqrt{L}   M_{L,L+1}(q\vec{r}_i)
                +\sqrt{L+1} M_{L,L-1}(q\vec{r}_i)
                 \Bigr], \nonumber \\
 {\cal L}^5_L(q)
         &=& \frac{1}{\sqrt{2L+1}}\sum_i
                  \Bigl( \frac{a_0}{2} +
                     \frac{a_1 m^2_\pi \tau^i_3}{2(q^2+m_\pi^2)}
                  \Bigr) \nonumber\\
         && ~~~~~~~~~~~~~~~                 
                  \Bigl[
                 \sqrt{L+1} M_{L,L+1}(q\vec{r}_i)
		+\sqrt{L}   M_{L,L-1}(q\vec{r}_i)
                 \Bigr], \nonumber \\
{\cal C}_L(q) 
	&=& \sum_{i,\ {\rm nucleons}} C_0^E j_L(qr_i)Y_L(\hat{r}_i), ~~~~
{\cal C}_0(q) = \sum_{i} C_0^E j_0(qr_i)Y_0(\hat{r}_i),  
\end{eqnarray} 
       where
$ M_{L,L'}(q\vec{r}_i) = j_{L'}(qr_i)[Y_{L'}(\hat{r}_i)\vec{\sigma}_i]^L.$
	In the limit of zero momentum transfer $S^A_{\rm SD}(q)$ 
	reduces to
\begin{eqnarray}
S^A_{\rm SD}(0) &=& {1\over{4 \pi}} \vert\langle N \vert\vert\sum_i
{1\over{2}}(a_0 + a_1 \tau_3^i) {\bf \sigma}_i\vert\vert N
\rangle\vert^2 \nonumber\\
    &=& {1\over{4 \pi}} \vert (a_0 + a_1)\langle N \vert\vert
{\bf S}_p\vert\vert N\rangle + (a_0 - a_1)\langle N \vert\vert
{\bf S}_n\vert\vert N\rangle\vert^2 \\
&=& {1\over{\pi}} \frac{(2J+1)(J+1)}{J} \vert a_p \langle N \vert
{\bf S}_p\vert N\rangle + a_n \langle N \vert{\bf S}_n\vert
N\rangle\vert^2 \nonumber\\
     &=& \frac{2J+1}{\pi} J(J+1) \Lambda^2,
\end{eqnarray}
with $ \displaystyle 
\Lambda = {{\langle N\vert a_p {\bf S}_p + a_n {\bf S}_n
\vert N \rangle}\over{J}} =
{{a_p \langle {\bf S}_p \rangle}\over{J}}  +
{{a_n \langle {\bf S}_n \rangle}\over{J}}.
$ 

In accordance with convention the Z components of the 
angular momentum and spin operators are evaluated in the
maximal $M_J$ state, e.g.
$\langle {\bf S} \rangle \equiv \langle N \vert {\bf S} \vert N \rangle
= \langle J,M_J = J \vert S_z \vert J,M_J = J \rangle$.

	In the ISPSM only the last odd nucleon contributes 
        to the spin and the angular momentum of the nucleus. 
	In this limit  
\begin{equation}
\langle{\bf S}^A_n\rangle
	= { J_A(J_A+1) - L_A(L_A+1) +{3 \over 4} \over 2J_A+2},
\end{equation}
      where $J_A$ and $L_A$ are the single-particle total and
      angular momenta. 
      They are deduced from the measured 
      nuclear angular momentum and the parity.  

\subsection{Nucleon spin structure}
	To evaluate the spin content of the nucleon
	one needs the matrix element of the 
	effective quark axial-vector current 
	$ J^\mu = \bar q\gamma^\mu\gamma_5 q $ \ in the nucleon
\cite{Jungman:1996df}.
	These matrix elements 
\begin{equation}
\label{Nucleon.spin.definition}
	\langle (p,n) | \bar{q} \gamma_\mu \gamma_5 q | (p,n) \rangle
		= 2 s_\mu^{(p,n)} \Delta q^{(p,n)}
\end{equation}
	are proportional to the spin of the neutron (proton or neutron),
	$s_\mu^{(p,n)}$.
	The quantities $\Delta q^{(p,n)}$ are usually extracted from
	the data obtained in polarized lepton-nucleon deep inelastic 
	scattering. 
	Uncertainties in the experimentally determined values
	for the quantities $\Delta q$ can lead to significant
	variations in the WIMP-nucleon axial-vector coupling,
	and therefore to the predicted rates for detection of
	WIMPs which have primarily spin couplings to nuclei
\cite{Jungman:1996df}.
    With definition
(\ref{Nucleon.spin.definition})
    the effective spin-dependent 
    interaction of neutralinos with the nucleon
    has the form 
\begin{equation}
\label{Nucleon.spin.neutralino-nucleon}
{\cal L}_{\rm spin} = 2 \bar{\chi} \gamma^\mu \gamma_5 \chi\; 
		\bar{n} s_\mu n
		\sum_{q=u,d,s}\, {\cal A}_q \,  \Delta q^{(n)}.
\end{equation}

      Recent global QCD analysis for the $g_1$ structure functions
\cite{Mallot:1999qb}, including ${\cal O}(\alpha_s^3)$ corrections,
	corresponds to the following values of spin nucleon parameters
\cite{Ellis:2000ds}
$$ 
\Delta_{u}^{(p)}\! = \Delta_{d}^{(n)}\! =   0.78 \pm 0.02, 	\ \ 
\Delta_{d}^{(p)}\! = \Delta_{u}^{(n)}\! =  -0.48 \pm 0.02,	\ \ 
\Delta_{s}^{(p)}\! = \Delta_{s}^{(n)}\! =  -0.15 \pm 0.02.
$$ 

\subsection{Effective neutralino-quark Lagrangian}
\setlength{\unitlength}{1mm}
 	The axial-vector and scalar 
	interaction of a neutralino with a
	quark $q$ is given by
$$
{\cal  L}_{eff} = 
	{\cal A}_{q}\cdot
      \bar\chi\gamma_\mu\gamma_5\chi\cdot
                \bar q\gamma^\mu\gamma_5 q +
	{\cal C}_{q}\cdot\bar\chi\chi\cdot\bar q q
	+	O (1/m_{\tilde q}^4). 
$$
       The terms with vector and pseudoscalar quark currents are
       omitted being negligible in the case of non-relativistic
       DM neutralinos with typical velocities $v_\chi\approx 10^{-3} c$.
	The Feynman diagrams which give rise to the effective 
	neutralino-quark axial-vector couplings 
\begin{eqnarray*}
{\cal A}_{q} =
	&-&\frac{g^{2}}{4M_{W}^{2}}
	   \left[
		\frac{{\cal N}_{14}^2-{\cal N}_{13}^2}{2}T_3 
		- \frac{M_{W}^2}{m^{2}_{\tilde{q}1} - (m_\chi + m_q)^2}
	   	(\cos^{2}\theta_{q}\ \phi_{qL}^2
	   + \sin^{2}\theta_{q}\ \phi_{qR}^2)
	\right. \\
        &-& \frac{M_{W}^2}{m^{2}_{\tilde{q}2} - (m_\chi + m_q)^2}
	     (\sin^{2}\theta_{q}\ \phi_{qL}^2
	     + \cos^{2}\theta_{q}\ \phi_{qR}^2) \\
        &-& \frac{m_{q}^{2}}{4}P_{q}^{2}
	\left(\frac{1}{m^{2}_{\tilde{q}1}
		- (m_\chi + m_q)^2}
             + \frac{1}{m^{2}_{\tilde{q}2}
                - (m_\chi + m_q)^2}\right) \\
        &-& \frac{m_{q}}{2}\  M_{W}\  P_{q}\  \sin2\theta_{q}\
            T_3 ({\cal N}_{12} - \tan\theta_W {\cal N}_{11}) \\
	&\times&
	\left.
\left( \frac{1}{m^{2}_{\tilde{q}1}- (m_\chi + m_q)^2}
    - \frac{1}{m^{2}_{\tilde{q}2} - (m_\chi + m_q)^2}\right)
	\right ] 
\end{eqnarray*}
	are shown in 
Fig.~\ref{Feynmans-spin-depnedent}.
\begin{figure}[t!]
\begin{picture}(100,25) 
\put(-5,-2){\includegraphics{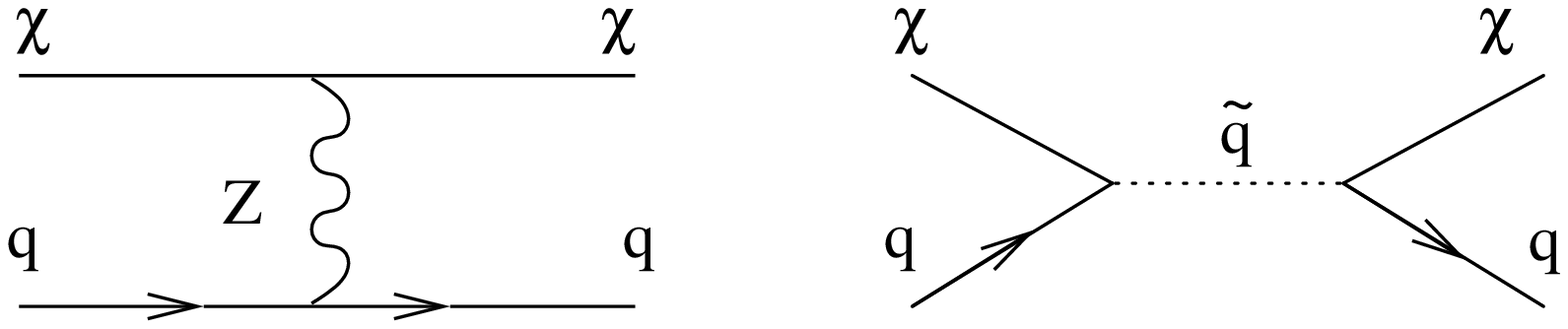}}
\end{picture}
\caption{Spin-dependent elastic scattering of neutralinos from quarks.}
\label{Feynmans-spin-depnedent}
\begin{picture}(100,25) 
\put(-5,-4){\includegraphics{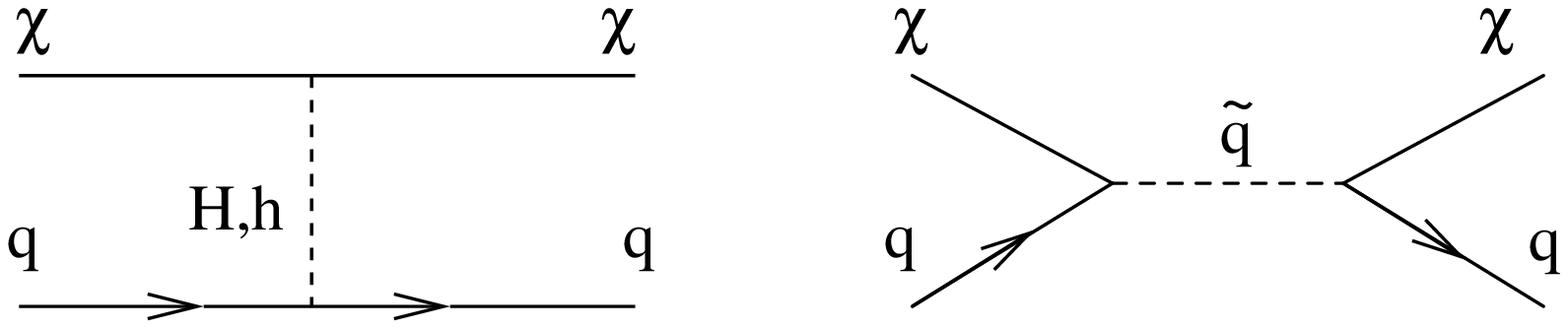}}
\end{picture}
\caption{Spin-independent or scalar (tree level) 
	elastic scattering of neutralinos from quarks.}
\label{Feynmans-spin-indepnedent}
\end{figure}
	The first term in ${\cal A}_{q}$
	comes from $Z^0$ exchange, 
	and the other terms come from squark exchanges.  
	The Feynman diagrams which give rise to the effective 
	neutralino-quark scalar couplings 
\begin{eqnarray*}
 {\cal C}_{q} =
	&-&  \frac{m_q}{M_{W}} \frac{g^2}{4} 
	\left[ 
	\frac{F_h}{m^2_{h}} h_q + \frac{F_H}{m^2_{H}} H_q 
	+ 
	\left(\frac{m_q}{4 M_W} P_{q}^{2} -
        \frac{M_W}{m_q} \phi_{qL}\ \phi_{qR}\right)
	\right.
\\&\times&
	\left(
	 \frac{\sin2\theta_{q}}{m^{2}_{\tilde{q}1} - (m_\chi + m_q)^2}
	-\frac{\sin2\theta_{q}}{m^{2}_{\tilde{q}2} - (m_\chi + m_q)^2}
	\right)
\\[3pt]
	&+&  \left. 
	P_q \left(
	\frac{\cos^{2}\theta_{q}\ \phi_{qL} -
         \sin^{2}\theta_{q}\ \phi_{qR}}{m^{2}_{\tilde{q}1} - (m_\chi + m_q)^2}
         -\frac{\cos^{2}\theta_{q}\ \phi_{qR} -
     \sin^{2}\theta_{q}\ \phi_{qL}}{m^{2}_{\tilde{q}2} - (m_\chi +
	m_q)^2}\right)
\right],\\[-25pt]
\end{eqnarray*}
where
\begin{eqnarray*}	
        F_{h} &=& ({\cal N}_{12} - {\cal N}_{11}\tan\theta_W)
        ({\cal N}_{14}\cos\alpha_H + {\cal N}_{13}\sin\alpha_H), \\
        F_{H} &=& ({\cal N}_{12} - {\cal N}_{11}\tan\theta_W)
        ({\cal N}_{14}\sin\alpha_H - {\cal N}_{13}\cos\alpha_H), \\
h_q &=&\bigl(\frac{1}{2}+T_3\bigr)\frac{\cos\alpha_H}{\sin\beta}
          - \bigl(\frac{1}{2}-T_3\bigr)\frac{\sin\alpha_H}{\cos\beta}, \\
H_q &=& \bigl(\frac{1}{2}+T_3\bigr)\frac{\sin\alpha_H}{\sin\beta}
      + \bigl(\frac{1}{2}-T_3\bigr)\frac{\cos\alpha_H}{\cos\beta},  \\
\phi_{qL} &=& {\cal N}_{12} T_3 + {\cal N}_{11}(Q -T_3)\tan\theta_{W},
\quad
\phi_{qR} = \tan\theta_{W}\  Q\  {\cal N}_{11},  \\
P_{q} &=&  \bigl(\frac{1}{2}+T_3\bigr) \frac{{\cal N}_{14}}{\sin\beta}
          + \bigl(\frac{1}{2}-T_3\bigr) \frac{{\cal N}_{13}}{\cos\beta},
\end{eqnarray*}
	are shown in 
Fig.~\ref{Feynmans-spin-indepnedent}.
	The importance of these scalar spin-independent contribution was
	found by K.Griest in
\cite{Griest:1988yr}.

\subsection{SUSY particle spectrum} 
     For completeness, we collect here formulas for masses of the SUSY
     particles in the MSSM. 
     There are four Higgs bosons --- 
     neutral $CP$-odd ($A$),  $CP$-even ($H,h$), charged ($H^{\pm}$).
     The $CP$-even Higgs boson mass matrix has the form:
\begin{eqnarray*}
         \left(\begin{array}{cc}
               H_{11} & H_{12}\\  
               H_{12} & H_{22} 
          \end{array}\right)
	&=&
	\half \left(\begin{array}{cc}
                 \tan\beta & -1 \\   -1 & \cot\beta 
         \end{array}\right) 
                  M_A^2\sin 2\beta                
\\ 
      &+&\half \left(\begin{array}{cc}
                  \cot\beta & -1 \\   -1 & \tan\beta 
           \end{array}\right) m_Z^2\sin 2\beta 
       +\omega
         \left(\begin{array}{cc}
               \Delta_{11} & \Delta_{12}\\  
               \Delta_{12} & \Delta_{22} 
          \end{array}\right),
\end{eqnarray*}
$$ H^{}_{11} = \frac{\sin 2\beta}{2}
        ( \frac{m^2_Z}{\tan\beta} + M^2_A\tan\beta )
                + \omega\Delta_{11},
$$
$$ H^{}_{22} = \frac{\sin 2\beta}{2}
       ( m^2_Z \tan\beta + \frac{M^2_A}{\tan\beta} )
                + \omega\Delta_{22},
$$
$$ H^{}_{12} = H^2_{21} = - \frac{\sin 2\beta}{2}
        ( m^2_Z + M^2_A )
                + \omega\Delta_{12}.
$$
	For example, $\Delta_{11}$ which includes loop corrections is
\begin{eqnarray*}
 \Delta_{11} &=& \frac{m^4_b}{c^2_\beta}
 (  \ln{\frac{m^2_{\tilde{b}_1} m^2_{\tilde{b}_2}} {m^4_b}}
  + \frac{2 {A}_b( {A}_b-\mu \tan\beta)}
         {m^2_{\tilde{b}_1}-m^2_{\tilde{b}_2}}
   \ln{ \frac{m^2_{\tilde{b}_1}} {m^2_{\tilde{b}_2}} }  ) \\
   &+& \frac{m^4_b}{c^2_\beta}
( \frac{ {A}_b( {A}_b-\mu \tan\beta)}
      {m^2_{\tilde{b}_1}-m^2_{\tilde{b}_2}})^2
 g(m^2_{\tilde{b}_1}, m^2_{\tilde{b}_2} ) 
 +\frac{m^4_t}{s^2_\beta}
 ( \frac{ \mu( {A}_t-\frac{\mu}{\tan\beta})} 
        {m^2_{\tilde{t}_1}-m^2_{\tilde{t}_2}})^2
 g(m^2_{\tilde{t}_1}, m^2_{\tilde{t}_2} ).
%
\end{eqnarray*}
	$\omega = \frac{3g_2^2}{16\pi^2m_W^2}$ 
	$c^2_\beta = \cos^2\!\beta$, $s^2_\beta = \sin^2\!\beta$, 
$g(m_1^2,m_2^2) =  2 - \frac{m_1^2+m_2^2}{m_1^2-m_2^2}
                       \ln{ \frac{m_1^2}{m_2^2} }.$
        The diagonalization of the above matrix gives the 
	Higgs boson masses $m_{H,h}$.
\begin{eqnarray*}
m_{H,h}^2
            &=&\half \Bigl\{ H_{11}+H_{22}
            \pm \sqrt{(H_{11}+H_{22})^2 - 4(H_{11}H_{22}-H_{12}^2)} \Bigr\},
\\
m^2_{H^\pm}&=&m^2_W + M^2_A   + \omega\Delta_{\rm ch}.
\end{eqnarray*}
        Here  $m_{H^\pm}$ is the charged Higgs boson mass in the one-loop
	approximation.
	The mixing angle $\alpha_H^{}$ is obtained from
$$
   \sin 2\alpha_H^{}= \frac{2H^2_{12}}{m^2_{H^0_1} - m^2_{H^0_2}},
\qquad
   \cos 2\alpha_H^{}= \frac{H^2_{11}-H^2_{22}}{m^2_{H^0_1} - m^2_{H^0_2}}.
$$ 
   The neutralino mass matrix in the basis  
   	($\tilde{B}$, $\tilde{W}^{3}$, $\tilde{H}_{1}^{0}$,
	 $\tilde{H}_{2}^{0}$) has the form:
$$
  {\cal M}_{\chi} = 
\left(
  \begin{array}{cccc}
    M_1 
	& 0   &-M_Z\cos\beta \sin\theta_W & M_Z\sin\beta \sin\theta_W  \\
    0   & M_2 & M_Z\cos\beta \cos\theta_W &-M_Z\sin\beta \cos\theta_W  \\
   -M_Z\cos\beta \sin\theta_W & M_Z\cos\beta \cos\theta_W & 0   & -\mu \\
    M_Z\sin\beta \sin\theta_W &-M_Z\sin\beta \cos\theta_W &-\mu & 0
  \end{array} \right). 
$$
	The diagonalization gives mass eigenstates (4 neutralinos): 
$$
\chi_i^{}(m_{{\chi}_i}) 
	= {\cal N}_{i1} \tilde{B} +  {\cal N}_{i2}  \tilde{W}^{3} +
{\cal N}_{i3} \tilde{H}_{1}^{0} + {\cal N}_{i4} \tilde{H}_{2}^{0}.
$$ 
        The lightest (LSP) $\chi = \chi_1$ is the best DM candidate.  
	The chargino  mass term is
$$
 \left(\tilde W^-, \tilde H_1^- \right) 
\left(\begin{array}{cc}
     M_2                  & \sqrt{2}M_W\sin\beta \\
     \sqrt{2}M_W\cos\beta & \mu
  \end{array} \right)
\left(  \begin{array}{c}
     \tilde W^+\\
     \tilde H_2^+ \end{array} \right)\ \  +  \ \ \mbox{h.c.}
$$
	The diagonalization
$   	U^* {\cal M}_{\tilde\chi^{\pm}} V^{\dagger} =
  	\mbox{diag}(M_{\tilde\chi^{\pm}_1}, M_{\tilde\chi^{\pm}_2})$
	gives charged mass eigenstates
$$
	\tilde\chi^{-} = U_{i1}\tilde W^- +  U_{i2}\tilde H^-,   \ \
	\tilde\chi^{+} = V_{i1}\tilde W^+ +  V_{i2}\tilde H^+
$$
	with masses
\begin{eqnarray*}
 M^2_{\tilde\chi^{\pm}_{1,2}} &=& \frac{1}{2}\left[M^2_2+\mu^2+2M^2_W
 \mp \right. \\
&\mp& \left.
  \sqrt{(M^2_2-\mu^2)^2+4M^4_W\cos^22\beta +4M^2_W(M^2_2+\mu^2+2M_2\mu
  \sin 2\beta )}\right].
\end{eqnarray*}
        The sfermion mass matrices ${\cal M}^2_{\tilde t}$, 
		       ${\cal M}^2_{\tilde b}$ 
		and ${\cal M}^2_{\tilde\tau}$ have the form:
\begin{eqnarray*}
{\cal M}^2_{\tilde t} &=&
\left[\begin{array}{cc}
m_{\tilde Q}^2+m_t^2+\frac{1}{6}(4M_W^2-M_Z^2)\cos 2\beta &
m_t(A_t -\mu\cot \beta ) \\
m_t(A_t -\mu\cot \beta ) &
m_{\tilde U}^2+m_t^2-\frac{2}{3}(M_W^2-M_Z^2)\cos 2\beta
\end{array}  \right],
\\[5pt]
{\cal M}^2_{\tilde b} &=&
\left[\begin{array}{cc}
m_{\tilde Q}^2+m_b^2-\frac{1}{6}(2M_W^2+M_Z^2)\cos 2\beta &
m_b(A_b -\mu\tan \beta ) \\
m_b(A_b -\mu\tan \beta ) &
m_{\tilde D}^2+m_b^2+\frac{1}{3}(M_W^2-M_Z^2)\cos 2\beta
\end{array}  \right],
\\[5pt]
{\cal M}^2_{\tilde\tau} &=&
\left[\begin{array}{cc}
m_{\tilde L}^2+m_{\tau}^2-\frac{1}{2}(2M_W^2-M_Z^2)\cos 2\beta &
m_{\tau}(A_{\tau} -\mu\tan \beta ) \\
m_{\tau}(A_{\tau} -\mu\tan \beta ) &
m_{\tilde E}^2+m_{\tau}^2+(M_W^2-M_Z^2)\cos 2\beta
\end{array}  \right].               
\end{eqnarray*}

     It is worth noting 
     that these masses as well as the above-mentioned
     couplings of neutralino-quark interactions 
     ${\cal A}_{\rm q}$ and ${\cal C}_{\rm q}$
     are functions of the common set of SUSY parameters 
     like, for example, $\tan\beta$, $M_A$, $\mu$,  $A_q$, etc.
     The set of parameters allows one to describe 
     observables at the 
     highest and lowerst energies 
     coherently and simultaneously. 


\bigskip 

{\small

\providecommand{\href}[2]{#2}\begingroup\raggedright\endgroup
}

\end{document}